\documentclass[traditabstract]{aa} % for the abstract without structuration 
\usepackage{graphicx}
\usepackage{color}
\usepackage{txfonts}
\usepackage{natbib}
\begin{document}
   \title{Sub-millimeter observations of {\it IRAS} and {\it WISE} debris disk candidates\thanks{{\it Herschel} is an ESA space observatory with science instruments provided by European-led Principal Investigator consortia and with important participation from NASA}}

 %  \subtitle{I. Overviewing the $\kappa$-mechanism}

   \author{J. Bulger
          \inst{1}
          \and
          T. Hufford\inst{2}
          \and
          A. Schneider\inst{2}
          \and
          J. Patience\inst{1,3}
          \and
          I. Song\inst{2}
          \and
          R. J. De Rosa\inst{3}
          \and
          A. Rajan\inst{3}
          \and
          C. D. Dowell\inst{4}
          \and
          D. McCarthy\inst{5}
          \and
          C. Kulesa\inst{5}
          }

\institute{School of Physics, University of Exeter, Exeter, EX4 6LZ, UK \\ \email{joanna@astro.ex.ac.uk}
\and
Department of Physics and Astronomy, The University of Georgia, Athens, GA USA 30602-2451
\and
School of Earth and Space Exploration, Arizona State University, Tempe, AZ USA 85281
\and
Department of Physics and Astronomy, Caltech, MC 249-17, Pasadena, CA 91125, USA
\and
The Department of Astronomy and Steward Observatory, The University of Arizona, Tucson, AZ USA 85721
}

   \date{Received April 15, 2013; accepted July 10, 2013}

% ################  ABSTRACT  ############################

% \abstract{}{}{}{}{} 
% 5 {} token are mandatory
 
  \abstract
  % context heading (optional)
  % {} leave it empty if necessary  
   { A set of six debris disk candidates identified with {\it IRAS} or {\it WISE} excesses were observed at either 350 $\mu$m or 450 $\mu$m with the Caltech Submillimeter Observatory. Five of the targets -- HIP 51658, HIP 68160, HIP 73512, HIP 76375, and HIP 112460 -- have among the largest measured excess emission from cold dust from {\it IRAS} in the 25-100 $\mu$m bands. Single temperature blackbody fits to the excess dust emission of these sources predict 350-450 $\mu$m fluxes above 240 mJy. The final  target -- HIP 73165 -- exhibits weak excess emission above the stellar photosphere from {\it WISE} measurements at 22 $\mu$m, indicative of a population of warm circumstellar dust. None of the six targets were detected, with 3$\sigma$ upper limits ranging from 51-239 mJy. These limits are significantly below the expected fluxes from SED fitting. Two potential causes of the null detections were explored -- companion stars and contamination. To investigate the possible influence of companion stars, imaging data were analyzed from new adaptive optics data from the ARIES instrument on the 6.5m MMT and archival {\it HST}, Gemini NIRI, and POSS/2MASS data. The images are sensitive to all stellar companions beyond a radius of 1-94 AU, with the inner limit depending on the distance and brightness of each target. One target is identified as a binary system, but with a separation too large to impact the disk. While the gravitational effects of a companion do not appear to provide an explanation for the submm upper limits, the majority of the {\it IRAS} excess targets show evidence for contaminating sources, based on investigation of higher resolution {\it WISE} and archival {\it Spitzer} and {\it Herschel} images. Finally, the exploratory submm measurements of the {\it WISE} excess source suggest that the hot dust present around these targets is not matched by a comparable population of colder, outer dust. More extensive and more sensitive {\it Herschel} observations of {\it WISE} excess sources will build upon this initial example to further define the characteristics of warm debris disks sources. }


\maketitle
%

% ################ TABLE 1: SAMPLE PROPERTIES ######################
\begin{table*}
\caption{Observed Sample}             
\label{table:Sample}      
\centering          
\begin{tabular}{l c c c c c c c c c c}   
\hline     
\noalign{\smallskip}     
Name & RA & Dec & Prop. mot. & D & SpTy & Age & Excess & T$_{\mathrm{dust}}$ & L$_{\mathrm{dust}}$ / L$_\star$ & Ref.  \\ 
& (J2000) & (J2000) & (\arcsec/yr) & (pc) & & (Myr) & source & (K) \\ 
\noalign{\smallskip}   
\hline\hline
\noalign{\smallskip}                       
HIP 51658 & 10:33:13.88 & +40:25:31.65 & -0.142, 0.008 & 34.6 $\pm$ 0.6 & A7 & 200 & {\it IRAS} & 40 & 1.06 $\times$ 10$^{\mathrm{-4}}$ & 1 \\
HIP 68160 & 13:57:16.13 & +23:21:44.37 & -0.330, -0.158 & 38 $\pm$ 1 & K0 & 230 & {\it IRAS} & & 1.49 $\times$ 10$^{\mathrm{-3}}$ & 2 \\
HIP 73165 & 14:57:11.01 & -04:20:47.32 & -0.097, -0.153 & 26.9 $\pm$ 0.2 & F0 & & {\it WISE} \\
HIP 73512 & 15:01:29.96 & +15:52:08.45 & 0.102, -0.238 & 30 $\pm$ 1 & K2 & 3000? & {\it IRAS} & 85 & 1.17 $\times$ 10$^{\mathrm{-3}}$ & 1\\
HIP 76375$^{\mathrm{a}}$ & 15:35:56.61 & +40:25:31.65 & -0.448, 0.051 & 22 $\pm$ 0.3 & K3 & 5000? & {\it IRAS} & 29 & 7.86 $\times$ 10$^{\mathrm{-4}}$ & 1\\
HIP 112460 & 22:46:49.81 & +44:20:03.10 & -0.705, -0.461 & 5.12 $\pm$ 0.05 & M3.5 & 500 & {\it IRAS} & 19 & & 3 \\
\noalign{\smallskip} 
\hline                  
\end{tabular}
\tablefoot{
\tablefoottext{a}{Known binary reported in the literature}
}
\tablebib{ 1. \citet{Rhee:2007}; 2. \citet{Moor:2006}; 3. \citet{Lestrade:2006} }
\end{table*}
% ################# END TABLE 1 ######################

% ################# SECTION: INTRO ######################
\section{Introduction}
Debris disks were first identified with the {\it Infrared Astronomical Satellite} ({\it IRAS}), from the observation that the far-IR flux of Vega \citep{Aumann:1984} was significantly brighter than expected from the stellar photosphere. Excess emission above the level of a stellar photosphere at wavelengths from the mid-IR to the millimeter provides evidence of reprocessed starlight emitted by a circumstellar debris disk of dust (c.f. reviews by \citealp{Zuckerman:2001, Wyatt:2008}). The disk origin of these excesses has been confirmed with spatially resolved imaging (e.g. \citealp{Smith:1984, Holland:1998}). Given the timescale for the dust to spiral into the star or be ejected from the system, these dust disks must be sustained by an ongoing process such as the collisional grinding of planetesimals into smaller particles \citep{Backman:1993} or an event such as a catastrophic collision of planets \citep{Cameron:1997}.

While {\it IRAS} provided an all-sky survey for debris disks with a typical sensitivity of (L$_{\mathrm{dust}}$ / L$_\star$) $\sim$ 10$^{\mathrm{-5}}$, subsequent pointed observations from the {\it Spitzer} Space Telescope (Werner et al. 2004), achieved an order of magnitude greater sensitivity. The combined results of the all-sky and pointed observations have identified over 200 debris disks among field stars (e.g. \citealp{Mannings:1998, Moor:2006, Rhee:2007, Habing:1999, Silverstone:2000, Rieke:2005, Su:2006, Bryden:2006, Beichman:2006, Hillenbrand:2008, Carpenter:2009}). Assuming a single temperature blackbody model for the excess emission in the spectral energy distribution, it is possible to classify debris disks as warm (equivalent to that of our Asteroid belt), with T$_{\mathrm{dust}}$ $\approx$100-250 K, that lie at distances $\la$10 AU, and cold debris disks (equivalent to that of our Kuiper belt), with T$_{\mathrm{dust}}$ $\la$50 K, typical distances of several hundred AU. 

With the more recent {\it Herschel} and Wide-field Infrared Survey Explorer {\it (WISE)} missions, it is possible to conduct pointed searches for cold debris disks down to the Kuiper belt level (L$_{\mathrm{dust}}$ / L$_\star$)  $\sim$10$^{\mathrm{-7}}$ to 10$^{\mathrm{-6}}$ \citep{Stern:1996} and all-sky surveys for warm debris disks at a sensitivity level that was previously only obtained with {\it Spitzer} pointed observations. In addition to the sensitivity offered by both {\it Herschel} and {\it WISE}, the significant improvement of spatial resolution is important to identify possible sources of contamination. Examples of debris disks identified through {\it IRAS} excesses (e.g. \citealp{Moor:2006, Rhee:2007}) that are actually due to confusion from background source contamination have been revealed by higher resolution ground-based observations carried out in the IR and submm (e.g. \citealp{Song:2002, Jayawardhana:2002, Lisse:2002, Sheret:2004}). \\ 

Mid-IR to far-IR observations have been critical in identifying the current population of debris disks, however measurements at longer wavelengths in the submm/mm wavelength range provide the best means to estimate one of the most important properties of a disk, the dust mass. For disks that are spatially resolved in the submm, it is also possible to determine additional disk characteristics such as  size and inclination which are degenerate with other disk parameters in an SED fit. For disks that have imaged asymmetries, these structures can encode the effects of gravitational interactions (e.g. \citealp{Liou:1999, Kuchner:2003, Wyatt:2006, Quillen:2006}). The submm/mm wavelength range is ideal for studying the interaction of the disk and planets, i.e. in the case of the multiple planet, debris disk system around HR$~$8799 (e.g. \citealp{Patience:2011, Hughes:2011}), since the large grains are expected to remain in resonances with the planets, while the smaller grains evolve into axisymmetric structures due to scattering or radiation pressure \citep{Wyatt:2006}. \\

In this paper, we present the results of a submm and IR imaging study of several {\it IRAS} excess stars and one {\it WISE} excess star. This sample includes previously known systems with cold dust \citep{Moor:2006, Rhee:2007} and newly identified debris disk candidates with evidence of either cold or warm dust. The properties of the sample are summarized in Section 2. The single-dish submm observations, infrared imaging and archival searches are described in Section 3. In Section 4, the data reduction and analysis of our submm and IR observations is presented. The results and implications of the IR image analysis is given in Section 5. In Section 6, we draw comparisons of our investigation to previous submm studies of debris disks. Finally, the conclusions of our study are given Section 7.

% ################## END: SECTION: INTRO  ###################### 

% ################ TABLE 2: SUBMM OBSERVATIONS ######################
\begin{table*}
\caption{Submm Observations}             
\label{table:SubmmObs}      
\centering          
\begin{tabular}{l c c c c c c}   
\hline     
\noalign{\smallskip}     
Name & UT Date & $\lambda$ ($\mu$m) & Exp. Time & $\tau$$_{\mathrm{225GHz}}$ range & Secondary Calibrators & Flux Calibrators \\ 
\noalign{\smallskip}   
\hline\hline
\noalign{\smallskip}                       
HIP 51658 & 2012 02 16 & 450 & 6$\times$10 min & 0.04-0.06 & 3C273, OH231.8, OJ287 & CRL618, OH231.8 \\
HIP 68160 & 2012 02 16 & 450 & 5$\times$10 min & 0.03-0.06 & 3C273, OJ287 & CRL618, OH231.8 \\
HIP 73165 & 2012 06 10 & 350 & 5$\times$10 min & 0.06-0.07 & Arp220 & Neptune, Uranus \\
HIP 73512 & 2012 06 10 & 450 & 6$\times$10 min & 0.06-0.07 & Arp220 & Arp220, Mars \\
HIP 76375 & 2012 06 10 & 350 & 8$\times$5 min & 0.05-0.06 & Arp220 & Neptune, Uranus \\
HIP 112460 & 2012 06 13 & 350 & 8$\times$10 min & 0.06-0.07 & Arp220 & Neptune, Uranus \\
\noalign{\smallskip} 
\hline                  
\end{tabular}
\end{table*}
% ################# END TABLE 2 ######################

% ################# SECTION: THE SAMPLE  ######################
\section{The sample}
The target sample for this investigation of cold and warm excess debris disks is drawn from two main sources -- the {\it IRAS} Faint Source Catalog (FSC v.2) \citep{Moshir:1992} and results from the {\it WISE} All-Sky Data Release \citep{Cutri:2012}. We selected a sample of cold excess disk candidates from the literature \citep{Lestrade:2006, Moor:2006, Rhee:2007} that satisfy three criteria: (1) a large {\it IRAS} excesses at 60 $\mu$m or 100 $\mu$m, (2) a distance within 100~pc, and (3) a declination above -30$\degr$. Spectral energy distributions (SEDs) were constructed, and a single temperature blackbody model was fit to the {\it IRAS} excesses. Targets with an expected 350 $\mu$m flux greater than 100~mJy were considered for this study, and the SEDs of the targets are shown in Figures \ref{fig:SEDcontam} and \ref{fig:SEDul}. Due to weather considerations, we focused our observations on sources with the greatest predicted 350~$\mu$m flux and and with nearby distances ($<$40~pc), in order to obtain meaningful limits and have the capacity to resolve the disks of the nearest systems. A total of 5 cold debris disks candidates were observed in this study, and are listed in Table \ref{table:Sample}, along with a summary of the modelled dust properties from the literature. The second type of target with evidence for warmer dust was drawn from the {\it WISE} (3.4 $\mu$m, 4.6 $\mu$m, 12 $\mu$m, and 22 $\mu$m) All-Sky Data Release. The identification of a  {\it WISE} W3 (12 $\mu$m) and/or W4 (22~$\mu$m) band excess was based on a careful cross-correlation of the {\it WISE} and {\it Hipparcos} catalogues \citep{Perryman:1997, vanLeeuwen:2007} with a distance cut of 100 pc and an estimate of the photospheric flux from theoretical atmosphere models \citep{Hauschildt:1999}. Due to the limited clear weather windows during the observing runs, it was only possible to observe one warm excess source -- HIP~73615. HIP 73615 has an excess only in the {\it WISE} W4 band, and the excess above the photosphere is 34 times the uncertainty in the photometry. The properties of the target are listed in Table \ref{table:Sample}, and the SED for this warm debris disk candidate is plotted in Figure~\ref{fig:SEDwarm}.

% ################# END SECTION: THE SAMPLE  ######################

% ################# SECTION: OBSERVATIONS  ######################
\section{Observations}
\subsection{Submm Observations}
Observations of the six sources were made with the SHARCII bolometer array \citep{Dowell:2003} on the 10.4m Caltech Sub-millimeter Observatory (CSO) on Mauna Kea. SHARCII contains a 12 $\times$ 32 array of pop-up bolometers with a $>$90 \% filling factor over the field and provides background-limited 350~$\mu$m and 450 $\mu$m maps. For the closest systems, we obtained 350~$\mu$m maps to potentially resolve the disks. For the remaining targets we measured 450 $\mu$m maps to measure longer wavelength fluxes distinct from the {\it Herschel} bands. Table \ref{table:SubmmObs} which wavelengths were observed for each source and the exposure times. The field-of-view of the SHARCII maps was $\sim$2\arcmin $\times$ 0\arcmin.6, and the beam sizes at 350 $\mu$m and 450 $\mu$m are $\sim$8$\farcs$5 and $\sim$10$\farcs$0, respectively. The beam shape was stabilized by the active primary surface correction (\citealp{Leong:2006}), which corrects for dish surface imperfections and changes in the gravity vector as a function of elevation. 

The data were acquired during two observing runs: February 14-22 2012 and June 6-14 2012. There were no significant changes to the instrument over the course of the project. The conditions during the observations ranged from excellent $\tau$$_{\mathrm{225GHz}}$~=~0.03 to marginal $\tau$$_{\mathrm{225GHz}}$ = 0.07.The observing sequence consisted of a series of target scans bracketed by scans of planets and nearby secondary calibrators, serving as both absolute flux calibration and pointing calibration measurements. Table \ref{table:SubmmObs} lists the calibrators observed for each science target. Pointing calibration measurements were taken at least once per hour during the observing period and focus corrections were monitored throughout the night. Typical individual target scan times were 620s, while 120s was sufficient for the calibrators.  A total of 1 to 11 scans were taken on the science targets.

% ################ TABLE 3: IMAGING ######################
\begin{table*}
\caption{High Resolution Imaging Observations}             
\label{table:Imaging}      
\centering          
\begin{tabular}{l c c c c c c c c}   
\hline     
\noalign{\smallskip}
\multicolumn{5}{c}{This Work} & \multicolumn{4}{c}{Archival Data} \\
\noalign{\smallskip}   
\hline\hline
\noalign{\smallskip}       
Name & UT Date & Filter & Instrument/ & Exp. Time & UT Date & Filter & Instrument/ & Exp. Time \\ 
& & & Telescope & (s) & & & Telescope & (s) \\
\noalign{\smallskip}   
\hline
\noalign{\smallskip}                       
HIP 51658 & 2012 12 23 & K$_{\mathrm{s}}$ & ARIES/MMT & 600 & 2010 02 04 & K$_{\mathrm{p}}$ & AOBIR/CFHT & 440 \\
HIP 68160 & 2012 12 23 & K$_{\mathrm{s}}$ & ARIES/MMT & 620 & 2008 03 21 & F606W & WFPC2/{\it HST} & 500 \\
HIP 73165 & - & - & - & - & 2005 08 15 & F160W & NICMOS/{\it HST} & 1663 \\
HIP 73512 & 2012 12 23 & K$_{\mathrm{s}}$ & ARIES/MMT & 450 & 2008 03 09 & F160W & NICMOS/{\it HST} & 1120 \\
HIP 76375 & - & - & - & - & 2008 03 13 & F160W & NICMOS/{\it HST} & 1120 \\
HIP 112460 & - & - & - & - & 2002 07 24 & FeII & AOBIR/CFHT & 64 \\
\noalign{\smallskip} 
\hline                  
\end{tabular}
\end{table*}
% ################# END TABLE 3 ######################

% #################  FIGURES:  SED EXCESS ################# 
 \begin{figure*}
\centering
\includegraphics[scale=0.38]{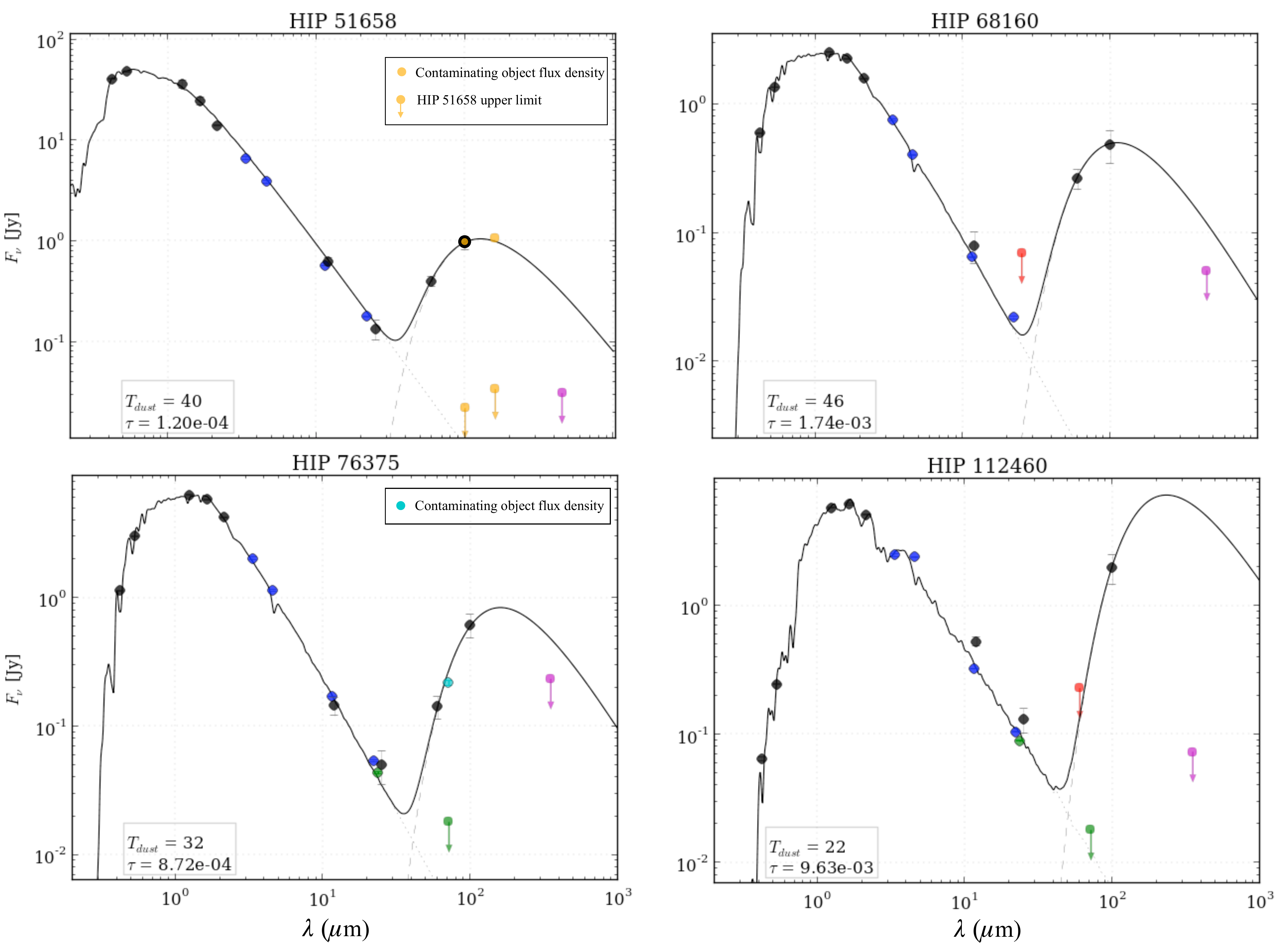}
\caption{SEDs of the cold debris disk source candidates for which the submm flux limits are substantially below the extension of the single temperature blackbody fit to the {\it IRAS} excesses -- HIP~51658, HIP~68160, HIP~76375 and HIP~112460. The NextGen atmospheric models corresponding to the best-fit stellar photosphere and the blackbody fit are plotted. The optical-near IR photometric points (black points) have been compiled from Tycho2 and 2MASS. The mid-IR photometric points have been drawn from {\it WISE} (dark blue points). For HIP~51568, with {\it Herschel} PACS photometry, the orange points with downward arrows correspond to the source 3$\sigma$ upper limits. The orange points represent the cumulative flux density of three nearby contaminating objects. For HIP~76375 and HIP~112460, with {\it Spitzer} MIPS photometry, the green points represent the source flux density and green points with downward arrows represent the source 3$\sigma$ upper limits. The flux density of the contaminating object measured in the MIPS 70 $\mu$m image for HIP 76375 is represented by the light-blue point. The CSO submm flux upper limit at 350 or 450 $\mu$m are represented by the pink points with downward arrows.
}
\label{fig:SEDcontam}
\end{figure*} 
% #################  END FIGURES: SED EXCESS ################# 

\subsection{Imaging Observations}
High angular resolution observations were used to search for the presence of a stellar binary companion to each member of the sample, since a companion can impact the stability of debris disk material. Dedicated observations were obtained for four targets with the Arizona Infrared Imager and Echelle Spectrograph (ARIES; \citealp{McCarthy:1998}) at the 6.5m AO-equipped MMT on December 23 2012. The observing strategy was the same for each target, with a series of short unsaturated exposures taken first to calibrate the photometry of the observing sequence, followed by deep exposures in which the core of the PSF of the target was saturated, increasing the sensitivity to faint stellar companions.  For both sets of exposures, images were obtained at a series of offset positions to measure the sky background and mitigate the effects of bad pixels. Additionally, the CFHT and {\it HST} archives were searched for previous high angular resolution observations of each sample member. Collectively, the new and archival observations include all the targets. Details of the high angular resolution observations of the sample are given in Table \ref{table:Imaging}.

To increase the range of separations at which a stellar binary companion can be detected to a limit of 10,000 AU, digitized photographic plates were obtained from the SuperCosmos Sky Survey Science Archive \citep{Hambly:2001} for each target within the sample. For the targets within 30 pc, several of the 15\arcmin $\times$ 15\arcmin digitized plates were combined to ensure the angular coverage was sufficient to detect co-moving companions with separations up to the 10,000 AU outer boundary considered for this study. The large time baseline between the first and last photographic observation, combined with the high proper motion of the sample members, allowed for an identification of wide common proper motion (CPM) companions. An example of one of the wide-field images is shown in Figure \ref{fig:Companion}.

\subsection{Archival mid-IR/far-IR images}
For each target, images from the all-sky surveys of {\it IRAS} and {\it WISE} were obtained for this program. Figure \ref{fig:IRAS} shows the region centered on each target for each of the four bands of {\it WISE} (left) and four bands of {\it IRAS} (right). Overplotted on each image is the approximate size of the 100 $\mu$m {\it IRAS} beam; the beam is estimated as circular using the average of the elliptical major and minor axis as the radius. We searched the {\it Spitzer} Heritage Archive and the {\it Herschel} Science Archive for higher resolution, pointed far-IR observations for each target, and three of the targets have additional higher resolution data. Level 2 MIPS data at 24 $\mu$m and 70 $\mu$m were retrieved for HIP 76375 and HIP 112460, and level 2.5 PACS data at 100 and 160~$\mu$m was retrieved for HIP 51658. The pointed observations with {\it Spitzer} and {\it Herschel} are shown in Figure \ref{fig:highIR}.

% ################# END SECTION: OBSERVATIONS  #####################

% ################# SECTION: DATA ANALYSIS  ###################### 
\section{Data Analysis}
\subsection{Submm maps and fluxes}
The analysis of the CSO 350 $\mu$m and 450 $\mu$m data included four main steps: application of a pointing model\footnote{http://www.submm.caltech.edu/$\sim$sharc/analysis/pmodel}, reduction of the raw data with the CRUSH pipeline \citep{Kovacs:2006,  Kovacs:2008}, measurement of aperture photometry, and calibration of fluxes. As an initial step, the pointing drifts of the telescope were estimated for each observation with a pointing model that accounts for both static effects, measured with many pointing sources at different positions throughout the run, and for time variable effects, measured with pointing sources observed before and after the science target. These offsets and the contemporaneous measurement of the atmospheric opacity are then incorporated into the data processing, implemented with the CRUSH (version 2.11-2) software \citep{Kovacs:2006,  Kovacs:2008}. The output of CRUSH includes an intensity map and signal-to-noise ratio map. The image processing algorithm is optimized depending on source
brightness. We utilized the data reduction settings appropriate to the flux levels for  the planets, pointing sources and science targets. Circular apertures with a radius of 9$\farcs$5 at 350 $\mu$m and 10$\farcs$6 at 450 $\mu$m were used for the target photometry; the aperture sizes are based on the measured FWHM of the point source calibrator at each wavelength.

Absolute fluxes of the planets are based on the distance from the Earth and the Sun, calculated at the time of observation and determined using the planetary brightness temperatures measured by \citet{Griffin:1993}. The planets Mars, Neptune and Uranus were used for primary flux calibration and have absolute calibration uncertainties of $\sim$5-10\% \citep{Griffin:1993}. For nights without a planet observation, the calibration is tied to the secondary calibrators Arp 220, CRL~618 and OH231.8. For observations at 350 $\mu$m, we use a flux density of 23.3 $\pm$ 4.2 Jy for CRL~618 (derived by C. D. Dowell and listed on the SHARCII website\footnote{http://www.submm.caltech.edu/$\sim$sharc/}). The uncertainty is based on the long term variability of evolved stars \citep{Sandell:2003, Jenness:2002}. At 450 $\mu$m, we use a flux density of 6.3 $\pm$ 0.8 mJy for Arp220 \citep{Dunne:2001}, 12.1 $\pm$ 2.2 mJy for CLR 618 and 12.7 $\pm$ 2.2 mJy for OH231.8 \citep{Jenness:2002}. The uncertainty of the flux calibration is combined with the uncertainty of the source signal which is measured from the CRUSH source maps. The total uncertainty on the submm fluxes of the science targets ranges from $\sim$15-30\%.

None of the six targets were detected in the CSO 350 $\mu$m or 450 $\mu$m maps. We have therefore pursued two types of analysis in order to investigate the causes for these null detections, presented in Section 5. With infrared imaging, we investigate the possible disruptive effects caused by companion stars and with an analysis of available mid-IR to far-IR maps, we explore the possibility of source contamination in the {\it IRAS} catalogue resulting in a misidentification of the target as a debris disk system.

% #################  FIGURES:  SED HIP 73512 ################# 
 \begin{figure}
\centering
\includegraphics[scale=0.38]{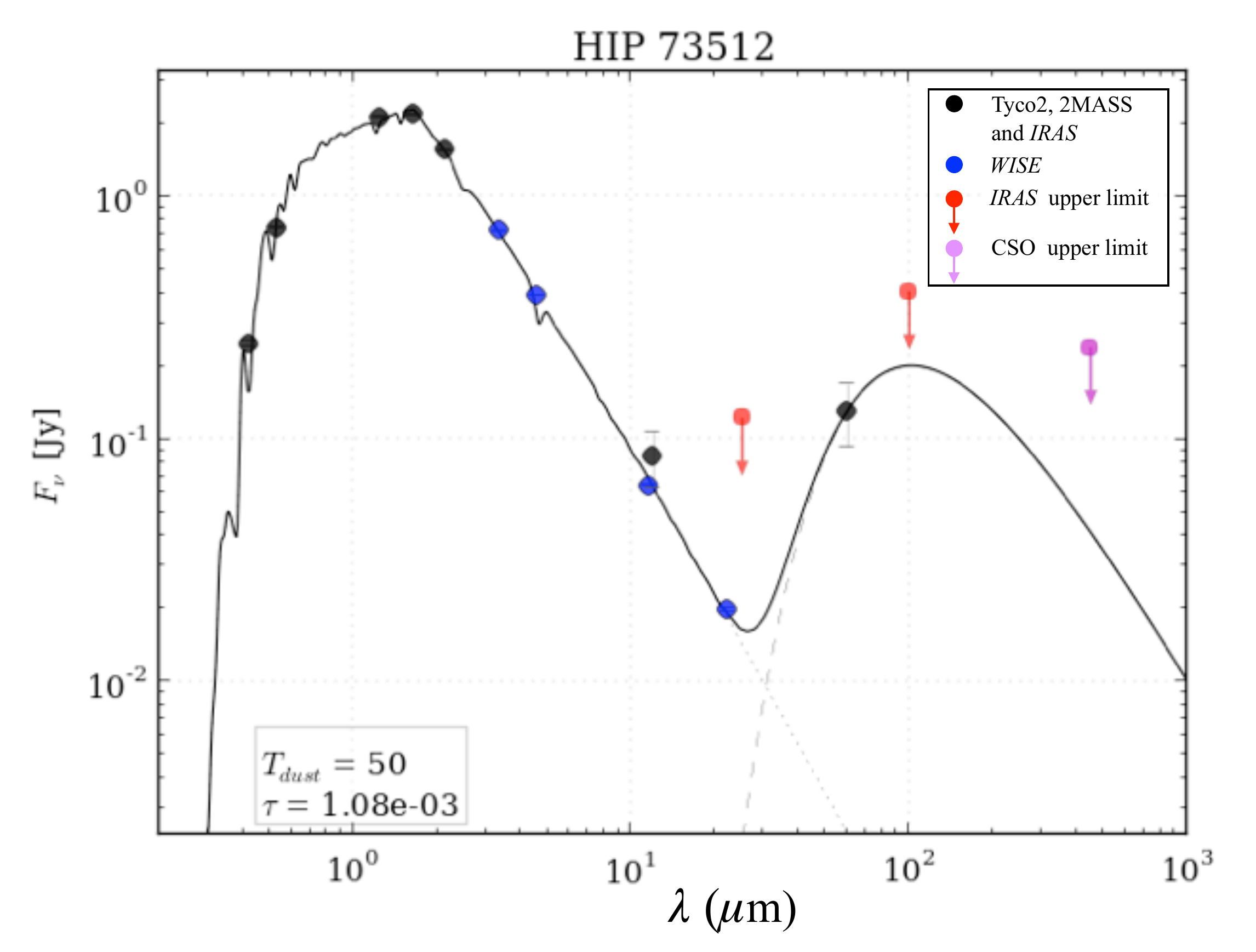}
\caption{SED of the cold debris disk candidate - HIP~73512 for which the submm flux limit lies above the  extension of the single temperature blackbody fit to the {\it IRAS} excesses. The NextGen atmospheric model corresponding to the best-fit stellar photosphere and the blackbody fit are plotted. Deeper submm observations for are required to further assess this source as a cold debris disk candidate.
}
\label{fig:SEDul}
\end{figure} 
% #################  END FIGURES: SED HIP 73512 ################# 

% #################  FIGURES:  SED HIP 73165 ################# 
 \begin{figure}
\centering
\includegraphics[scale=0.38]{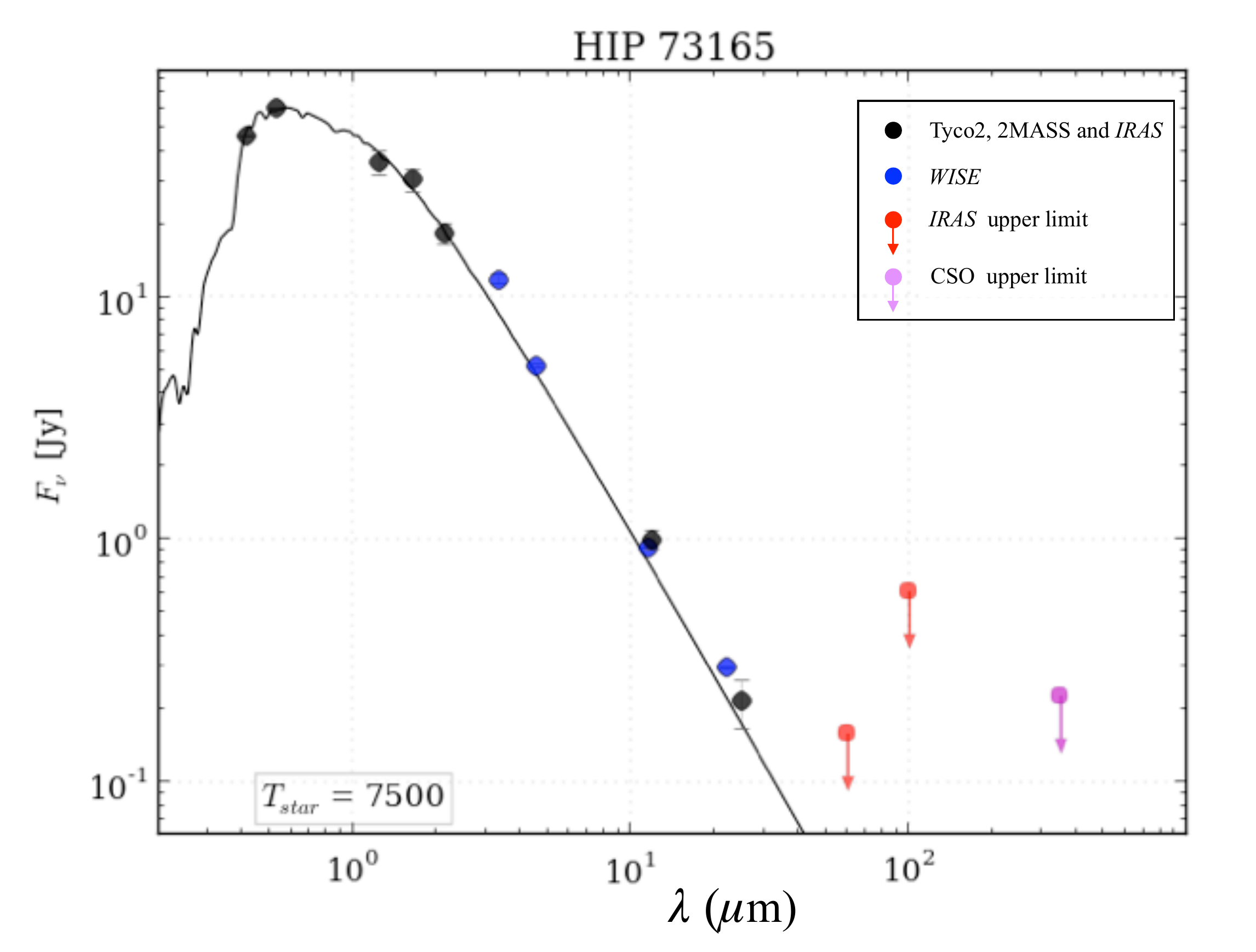}
\caption{SED of our {\it WISE} excess source - HIP~73165. The NextGen atmospheric model corresponding to the best-fit stellar photosphere is plotted. Excess is seen in the W4 band above the photosphere, and is 34 times the uncertainty in the photometry. The CSO 450 $\mu$m 3$\sigma$ upper limit (pink point and downwards arrow) shows the level to which lack of submm excess emission is seen.}
\label{fig:SEDwarm}
\end{figure} 
% #################  END FIGURES: SED HIP 73165 ################# 

% ################# FIGURE: HIP76375 COMPANION  #################
\begin{figure}
\centering
\includegraphics[scale=0.2]{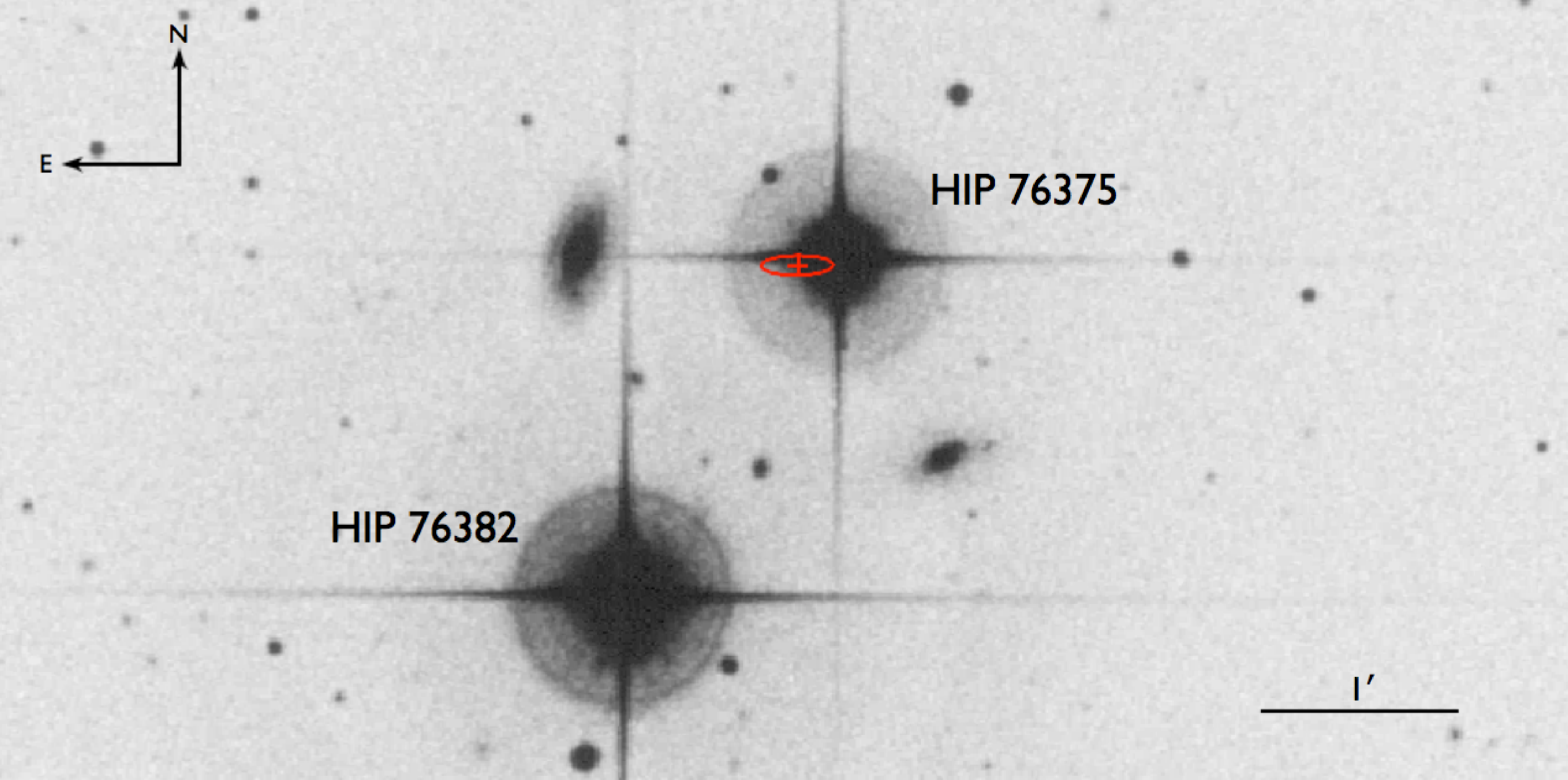}
\caption{Field plate image of HIP~76375, the only source in our sample with a known companion. The orientation is indicated with the arrows, along with a scale bar for this 8\arcmin $\times$ 4\arcmin image. The companion - HIP~76382, identified from CPM \citep{Shaya:2011} is labeled on the image and lies 148$\degr$ from the primary, with a separation of 121\farcs8. The {\it IRAS} FSC positional uncertainty for HIP~76375 is displayed on the image (red ellipse).
}
\label{fig:Companion}
\end{figure}
% ################# END FIGURE: HIP76375 COMPANION  #################

% ################ TABLE 4: DISK MASS ######################
\begin{table*}
\caption{ Sub-millimeter Detection Limits}             
\label{table:Diskmass}      
\centering          
\begin{tabular}{l c c c c}   
\hline     
\noalign{\smallskip}     
Target & $\lambda$ & Map noise rms & Flux limit$^{\mathrm{a}}$ & Dust mass limit$^{\mathrm{b}}$ \\ 
& ($\mu$m) & (mJy/beam) & (mJy) & (M$_{\oplus}$)  \\ 
\noalign{\smallskip}   
\hline\hline
\noalign{\smallskip}                       
HIP 51658 & 450 & 8 & $<$31 & $<$0.046 \\
HIP 68160 & 450 & 13 & $<$51 & $<$0.089 \\
HIP 73165 & 350 & 66 & $<$227  & $<$0.094 \\
HIP 73512 & 450 & 66 & $<$239 & $<$0.262 \\
HIP 76375 & 350 & 68 & $<$234 & $<$0.065 \\
HIP 112460 & 350 & 22 & $<$73 & $<$0.001 \\
\noalign{\smallskip} 
\hline                  
\end{tabular}
\tablefoot{
\tablefoottext{a}{Flux calibration uncertainties are included in the 3$\sigma$ limits reported.} \\
\tablefoottext{b}{Calculated assuming an average dust temperature of 30 K.}
}
\end{table*}
% ################# END TABLE 4 ######################

\subsection{High resolution and wide-field, near-IR/optical companion search images}
For the ground-based AO images from the MMT and CFHT, we performed standard image processing steps to account for sky background and detector dark current, nonlinearity, bad pixels, and quantum efficiency variations. The individual science images were aligned, and median combined. For the {\it HST} archive data, the images were already reduced and flux calibrated, so no additional processing was required. Each final target image was searched for stellar companions, and detection limits were calculated as a function of separation. For each target, the minimum separation at which a stellar companion at the bottom of the Main Sequence (0.08~M$\sun$) could be detected was determined. The variance of the pixel values within the science images was measured as a function of separation, scaled by the flux and apparent magnitude of the central target, and compared to the expected apparent magnitude of a 0.08 M$\sun$ stellar companion obtained from theoretical models \citep{Baraffe:1998}.

Three of the sample members have also been searched for binary companions with speckle interferometry (HIP~51658, HIP~76375, and HIP~112460; \citealp{Miura:1995, Hartkopf:2009, Balega:2007}), and none of which were found to have a binary companion. While these techniques would be sensitive to a high mass ratio binary companion at close separations, their sensitivity does not extend to the bottom of the Main Sequence. Future observations with increased sensitivity are required in order to rule out the presence of any stellar companion interior to the detection limits given in Table \ref{table:Companion}.

To search for companions beyond the field-of-view of the high angular resolution instruments, the digitized photographic plates for each target were blinked to reveal the presence of any wide co-moving companions. To ensure wide CPM companions were not missed during the visual inspection, the PPMXL \citep{Roeser:2010} and the UCAC4 \citep{Zacharias:2012} catalogues were also searched for stars with proper motions consistent with the object being physically bound to the primary.

\subsection{{\it IRAS} data}
We visually inspected each image in order to assess the quality and association of {\it IRAS} Faint Source Catalogue fluxes that have been extracted for the targets. The synthesized beam sizes of the Faint Source Survey images are approximately 1\arcmin $\times$ 4\arcmin at 12 $\mu$m and 25 $\mu$m, 2\arcmin $\times$ 4\arcmin and 4\arcmin $\times$ 5\arcmin at 60~$\mu$m and 100 $\mu$m, respectively. The positional uncertainties of {\it IRAS} detections for our sample is $\sim$3\arcsec in the in-scan direction and $\sim$18\arcsec in the cross-scan direction. In addition, we have reassessed the quality of the FSC extracted flux densities for our targets, with the use of the online Scan Processing and Integration ({\it SCANPI}) tool\footnote{http://irsa.ipac.caltech.edu/applications/Scanpi/}. {\it SCANPI} combines the signal for all (in-scan) survey scans that correspond to the target position and provides fluxes of faint sources, estimates of true local upper limits and deviation of the signal peak to that of the source position. Despite the stringent processing of the survey scans and catalog criteria that have been implemented for source extraction and identification, the large angular resolution of the {\it IRAS} beams is one of the primary factors of possible source contamination. 

\subsection{Higher resolution mid-far IR {\it WISE}, {\it Spitzer} and {\it Herschel} images}
{\it WISE} provides a factor of 10 higher resolution images in the mid-IR than {\it IRAS}. The {\it WISE} beam sizes are 6$\farcs$1, 6$\farcs$4, 6$\farcs$5 and 12\arcsec at 3.4 $\mu$m (W1), 4.6 $\mu$m (W2), 12 $\mu$m (W3) and 22 $\mu$m (W4). To utilize this significant improvement of resolution, we performed visual cross inspection of the four {\it WISE} bands, to identify any contaminating objects that are contained within the {\it IRAS} beam sizes. Additionally, we investigated the available {\it Spitzer} MIPS and {\it Herschel} PACS \citep{Poglitsch:2010} processed images from the archive for the targets HIP 76375, HIP 112460, and HIP 51658. HIP 68160 was not observed by either {\it Spitzer} or {\it Herschel}. The {\it Spitzer} FWHM beam size with MIPS at 24 $\mu$m is 6\arcsec, a factor two greater than the {\it WISE} W4 band, and pointed observations further increase the sensitivity beyond that achieved with the {\it WISE} survey. The far-IR MIPS channels at 70 $\mu$m and 160 $\mu$m, with corresponding beam sizes of 20$\arcsec$ and 40$\arcsec$, and the {\it Herschel} PACS channels at 70~$\mu$m , 100~$\mu$m and 160~$\mu$m, with FWHM beam sizes of 5$\farcs$6, 6$\farcs$8 and 11$\farcs$4 respectively, provide the far-IR wavelength coverage with significantly greater resolution than that of {\it IRAS}, enabling the identification of contaminating objects such as background galaxies that become increasingly brighter across the mid-far IR regime.

% ################# END SECTION: DATA ANALYSIS  ###################### 

% ################ TABLE 5: COMPANION RESULTS ######################
\begin{table*}
\caption{Companion Search Results}             
\label{table:Companion}      
\centering          
\begin{tabular}{l c c c c c c}   
\hline     
\noalign{\smallskip}     
Target & Radius to reach bottom of MS & Detected & Companion Separation & Position Angle & $\Delta$mag & Technique \\ 
& (AU) & Companion & (AU) & (\degr) & (mag)  \\ 
\noalign{\smallskip}   
\hline\hline
\noalign{\smallskip}                       
HIP 51658 & 39 & N & - & - & - & - \\
HIP 68160 & 94 & N & - & - & - & - \\
HIP 73165 & 45 & N & - & - & - & - \\
HIP 73512 & 41 & N & - & - & - & - \\
HIP 76375 & 35 & Y & 2725 & 148 & $\Delta$K=0.79 & CPM \\
HIP 112460 & 1 & N & - & - & - & - \\
\noalign{\smallskip} 
\hline                  
\end{tabular}
\end{table*}
% ################# END TABLE 5 ######################

% ################# SECTION: RESULTS ################# 
\section{Results}
None of the six debris disk targets were detected in the CSO SHARCII maps. The 3$\sigma$ upper limits range from 31 mJy for HIP 51658 to 239 mJy for HIP 73512 at 450~$\mu$m, and all the limits are reported in Table \ref{table:Diskmass}. For four of the five cold debris disks targets, the submm flux limits are substantially below the extension of the blackbody fit to the {\it IRAS} excesses, as shown in Figure \ref{fig:SEDcontam}. The CSO limit for the cold debris disk candidate HIP 73512 is above the blackbody fit, as shown in Figure \ref{fig:SEDul}, and the CSO data do not constrain the disk in this case. The {\it WISE} excess source HIP 73165 also shows a lack of submm emission at a level reported in Table \ref{table:Diskmass} and plotted on the SED in Figure \ref{fig:SEDwarm}. 

\subsection{CSO flux limits}
The flux limits were converted into dust mass limits following previous submm studies (e.g. \citealp{Zuckerman:1993}) using the expression $M_{\mathrm{d}}$ = ($F_{\nu}D^{2})/B_{\nu}(T){\kappa}_{\nu}$, where the blackbody function $B_{\nu}(T)$ is equal to 2$kT/{\lambda}^{2}$ in the Rayleigh-Jeans limit. The dust opacity at the frequency of observation ${\kappa}_{\nu}$ assumes a functional form ${\kappa}_{\nu} = {\kappa}_{0}({\nu} / {\nu}_{0})^{~\beta}$ at submm wavelengths. We use a normalization value ${\kappa}_{0} = 0.17~\mathrm{m}^{2}~\mathrm{kg}^{-1}$ at 850~$\mu$m and scale with $\beta$=1 \citep{Zuckerman:1993,Pollack:1994}. Existing submm studies find that $\beta$=1 is a representative value of debris disks \citep{Dent:2000} and the normalization value for dust opacity is adopted for direct comparison with previous studies (e.g \citealp{Sylvester:2001, Wyatt:2003, Liu:2004, Sheret:2004, Najita:2005, Williams:2006}). Assuming an average dust temperature of 30 K for the submm-emitting dust, the submm flux limits correspond to dust mass limits of 0.001 to 0.261 M$_{\oplus}$, as reported in Table \ref{table:Diskmass}. These dust mass limits are comparable to the detections of other nearby debris disks based on either submm fluxes or fits to {\it IRAS} fluxes over a large wavelength range (e.g. compilation in \citealp{Rhee:2007}).

The submm upper limits are plotted along with previous measurements at shorter wavelengths to construct the SEDs of the {\it IRAS} cold excess debris disks with limits below the blackbody fit in Figure \ref{fig:SEDcontam} and with a limit above the blackbody fit in Figure \ref{fig:SEDul}. The {\it WISE} warm excess debris disk SED is plotted in Figure \ref{fig:SEDwarm}. For the cold excess sources, the fits to the 60-100 $\mu$m emission, assuming a single temperature blackbody, are also shown for comparison with the measured CSO limits. For the targets HIP 51658, HIP 68160, HIP 76375 and HIP 112460, the submm limits are all well below the expected flux level at these wavelengths, making these limits more restrictive than previous submm measurements of some debris disks candidates \citet{Williams:2006}. For the {\it WISE} source, there is not a large enough wavelength coverage for the excess emission to perform a fit, and the exploratory measurement at the longer wavelength is shown in comparison to the newly reported mid-IR excess in the {\it WISE} W4 band. With only an excess in the longest wavelength {\it WISE} band, it is not possible to fit a blackbody to estimate an expected flux level for the CSO measurement, but the non-detection is consistent with a spatially confined distribution of hot dust for this source. Larger scale studies with more sensitive limits from {\it Herschel} will provide more definitive constraints on the origin of {\it WISE} excesses for sources such as HIP 73165.

\subsection{Companion systems}
Since a companion star will gravitationally truncate the outer portion of a disk and reduce the amount of submm-emitting dust, all targets were searched for stellar companions. A visual inspection of the high-resolution data did not reveal any  companion candidates. Separately, a visual inspection of each of the blinked image pairs combined with a search of the astrometric catalogues, identified one co-moving object - the known wide common proper motion (CPM) companion to HIP 76375 \citep{Shaya:2011}. The HIP 76375 system is shown in Figure \ref{fig:Companion}. The binary separation is 121$\farcs$8, which corresponds to a projected physical separation of $\sim$2700 AU -- too large to impact the disk size unless the orbit is very eccentric and periastron passage occurs at a much smaller separation. Table \ref{table:Companion} lists the details of the HIP 76375 binary system and the closest separation at which the data become sensitive to companions at the bottom of the Main Sequence for each of the targets. The minimum separation ranges from only 1 AU for the M-star HIP~112460 at $\sim$5 pc to 94 AU for the more massive and more distant A-star HIP 51658 at $\sim$30 pc. In addition to the the companion search we performed, five targets were included in the interferometric \citep{vanBelle:2008, vanBelle:2010} and speckle \citep{Hartkopf:1984, Miura:1995, Hartkopf:2009, Balega:2007} searches, and no companions were discovered for these targets, although the detection limits were not deep enough to include all possible stellar companions.

% ################# IRAS & WISE IMAGES  #################
\begin{figure*}
\centering
\includegraphics[scale=0.45]{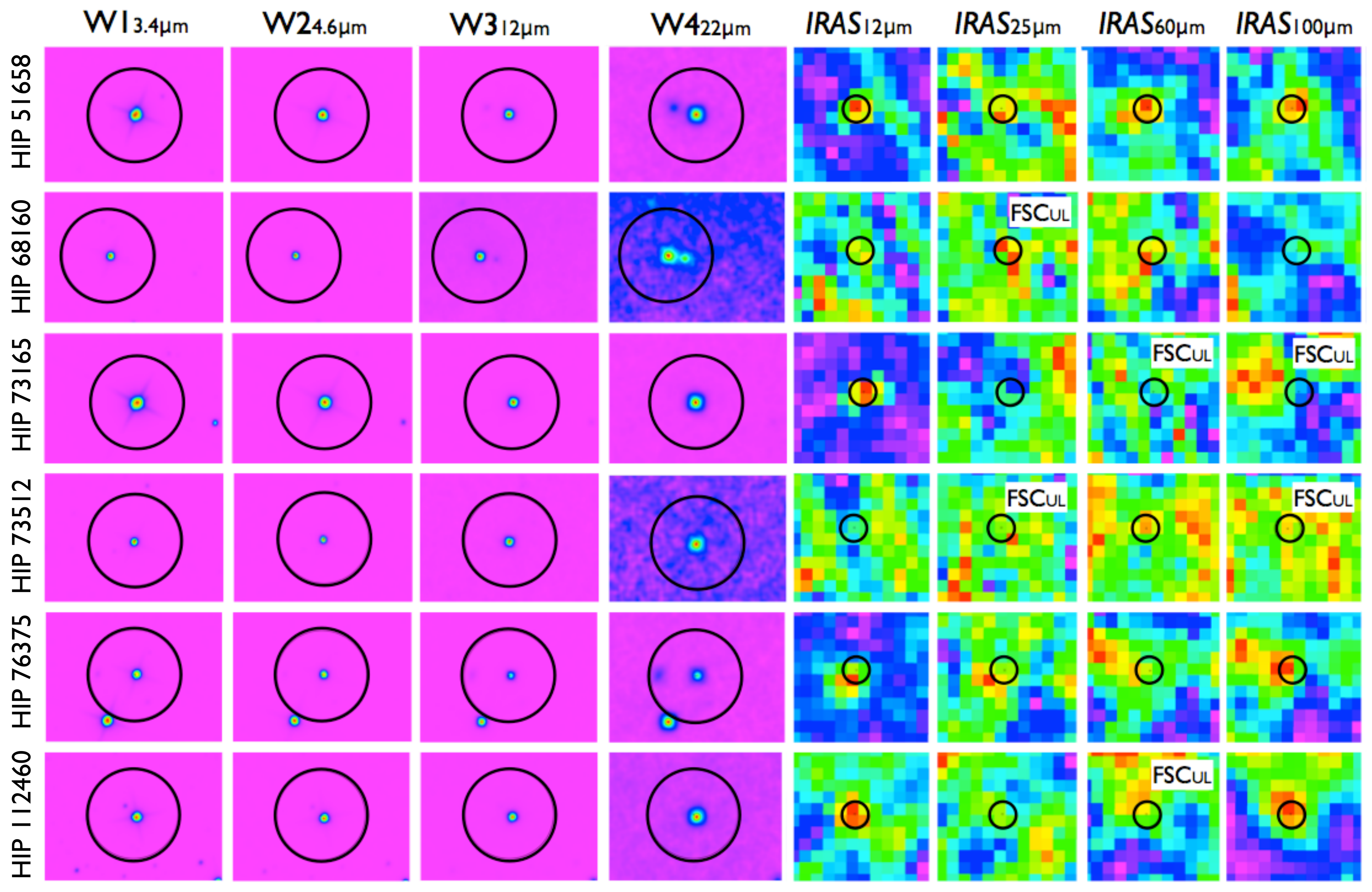}
\caption{The {\it WISE} All-Sky Survey Atlas images and {\it IRAS} Sky Survey Atlas (ISSA) for all the sources of our sample. The images are displayed per source (repented by the row heading) in order of survey, and the wavelength of observation from left to right (indicated by the column heading). Orientation for all images are North, up and East, left. The {\it WISE} images are 400\arcsec $\times$ 300\arcsec in size. The spatial resolution of the ISSA images at each wavelength are smoothed to the {\it IRAS} 100 $\mu$m beam size and are 900\arcsec $\times$ 900\arcsec in size. The black circles represent the {\it IRAS} 100~$\mu$m beam centered over the source positions. The sources and the wavelengths for which {\it IRAS} FSC upper limits are reported are indicted in the images. The four cold debris disks candidates with submm limits well below the SED fit are HIP~51658, HIP~61680 and HIP~76357 and HIP~112460, and are investigated for the possibility of contaminating sources within the {\it IRAS} beam. For HIP~51658, HIP~61680 and HIP~76357 nearby objects are visible in the higher resolution  {\it WISE} images (see also Figure \ref{fig:highIR}). Additionally, for HIP~76357, the secondary companion (HIP~76382) is seen dominant in all {\it WISE} images and is noticeably seen in the 12 $\mu$m and 25 $\mu$m ISSA images. No contaminating sources are visible in either the {\it WISE} or ISSA images for HIP~112460.}
\label{fig:IRAS}
\end{figure*}
% ################# END IRAS & WISE IMAGES  #################

\subsection{{\it IRAS} source contamination}
For the four cold debris disk candidates with submm limits well below the SED fit (shown in Figure \ref{fig:SEDcontam}), we investigated the possibility of contaminating sources within the {\it IRAS} beam. All targets have {\it WISE} images in four bands with higher resolution than {\it IRAS}, and three of the four targets have {\it Spitzer} data in at least one bandpass. Three of the four objects -- HIP 51658, HIP 61680, and HIP 76375 -- show contaminating sources in the higher resolution mid-IR/far-IR images shown in Figure \ref{fig:highIR}. Only one -- HIP 112460 -- shows a single object in the higher resolution {\it WISE} and {\it Spitzer} maps shown in Figure \ref{fig:hip112460}. The results of the contamination analysis for these sources is discussed in the following subsections.

\subsubsection{HIP 51658}
From the sequence of {\it WISE} images for HIP 51658 in Figures \ref{fig:IRAS} and \ref{fig:highIR}, a nearby object $\sim$55\arcsec to the NE of HIP~51658 is seen and becomes increasingly brighter with wavelength. Also shown in Figure \ref{fig:highIR} is the {\it Herschel} PACS 100 $\mu$m map centered on the target which reveals the offset source, but not the target. In both the {\it WISE} and {\it Herschel} maps there are additional fainter sources within the {\it IRAS} beam (not visible with the scaling of Figure \ref{fig:highIR}). Three distinct sources are detected in the {\it Herschel} 100 $\mu$m map. From our {\it SCANPI} analysis of this source, we find that signals are significant in all IRAS bands ($>$5$\sigma$) and the the deviation from source position to the signal peak is $<$0.5\arcmin.

To determine the flux of the contaminating objects, we have measured the flux densities of the nearby three objects that fall within the {\it IRAS} beam centered over the source position, and find that the cumulative flux density at 100 $\mu$m is consistent with the {\it IRAS} FSC 100 $\mu$m flux. Furthermore, the 160~$\mu$m cumulative flux density fits well to the single temperature blackbody, modelled excess fit. The PACS fluxes, along with our 450 $\mu$m upper limit that lies significantly below the fit, are shown in Figure \ref{fig:SEDcontam}. This analysis indicates that the associated {\it IRAS} 60~$\mu$m and 100~$\mu$m fluxes of this source are contaminated and that HIP 51658 is not a bona fide debris disk system. To determine more accurate flux upper limits for HIP 51658, we have extracted level 2.5 processed PACS 100 $\mu$m and 160~$\mu$m images from the {\it Herschel} archive and measured 3$\sigma$ flux upper limits of 22 mJy at 100~$\mu$m and 34~mJy at 160~$\mu$m, following the point source photometry guidelines recommended by the {\it Herschel} team\footnote{Technical Note PICC-ME-TN-037 in http://herschel.esac.esa.int}.

\subsubsection{HIP 68160}
As shown in Figure \ref{fig:highIR}, a nearby object $\sim$35\arcsec to the west of HIP~68160 is visible in the W4 22 $\mu$m image, and lies within the {\it IRAS} beam. The {\it IRAS} 25 $\mu$m flux for this source is reported as an upper limit in the FSC. As there are no higher resolution far-IR observations available from {\it Spitzer} or {\it Herschel} for this source, we are unable to quantify the level of flux contamination within the {\it IRAS} 60 $\mu$m and 100 $\mu$m bands, unlike the case of HIP 51658. The presence of the additional source in the {\it WISE} W4 image does, however, suggest that the {\it IRAS} fluxes are heavily influenced by this red object. Based on this second source in the field, HIP 68160 is also a probable case of a misclassified debris disk system.

\subsubsection{HIP 76375}
\begin{figure}
\centering
\includegraphics[scale=0.14]{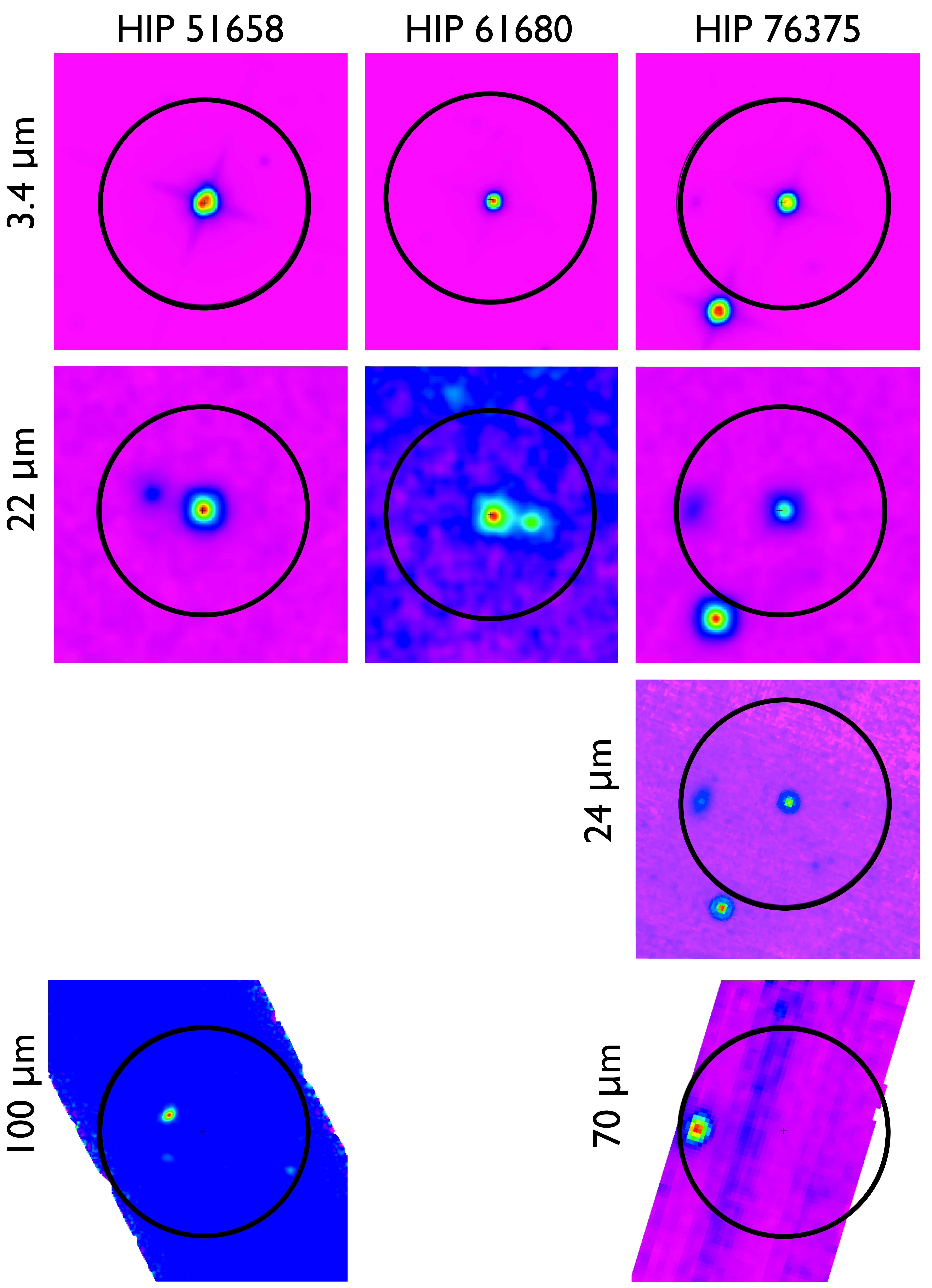}
\caption{Higher resolution IR images for the sources -- HIP~51568, HIP~61680 and HIP~76375, in which nearby contaminating objects are seen in the {\it WISE} and when available {\it Spitzer} MIPS and {\it Herschel} PACS images. The orientation for all images are North, up and East, left and the images are 300\arcsec in size along each axis. The black circles represent the {\it IRAS} 100~$\mu$m beam centered over the source positions. For HIP~51658 a nearby object ($\sim$55$\arcsec$ to the NE) is seen to become increasingly brighter at increasing wavelengths, as shown  in the {\it WISE} 22 $\mu$m (W4) and {\it Herschel} PACS 100~$\mu$m images. For HIP~68160, a nearby object ($\sim$35$\arcsec$ to the west) is seen in the W4 image. For HIP~76375 and the binary companion - HIP~76382 (121\arcsec to the SE) are both visible in the {\it WISE} 3.4 $\mu$m (W1), W4 and {\it Spitzer} MIPS 24 $\mu$m images. Neither component of the system are seen in the {\it Spitzer} MIPS 70 $\mu$m image. The galaxy IC~4563 ($\sim$80$\arcsec$ to the east), is visible in the W4 and MIPS images, and lies within the {\it IRAS} beam area.}
\label{fig:highIR}
\end{figure}

This target is the only known binary in our sample. Both the target and binary companion (separation of 121$\farcs$8) are visible in all four {\it WISE} images (see Figure \ref{fig:IRAS}) and the MIPS 24~$\mu$m image (see Figure \ref{fig:highIR}). From our {\it SCANPI} analysis we note that there is a large positional offset (1.9\arcmin) from the source to signal peak in the {\it IRAS} 12~$\mu$m band, with a SNR of 8. Figure \ref{fig:IRAS} shows the {\it IRAS} scans at 12~$\mu$m, 60~$\mu$m and 100~$\mu$m. In the 12 $\mu$m scan, the position of the binary companion falls on the peak emission pixel, indicating that the {\it IRAS} 12 $\mu$m flux reported for this source contaminated is from emission of the secondary. 

To investigate the longer wavelength measurements, we have extracted the archival MIPS images at 24~$\mu$m and 70~$\mu$m for this source, shown in Figure \ref{fig:highIR}. In all the {\it WISE} images and in the MIPS 24 $\mu$m image, both the target and companion are visible. The galaxy, IC~45634\footnote{Identified in the NASA/IPAC Extragalactic Database - http://ned.ipac.caltech.edu/} ($\sim$80\arcsec to the E) is seen to become increasingly brighter with longer wavelengths, as seen in Figures \ref{fig:IRAS} and \ref{fig:highIR}. In the MIPS 70 $\mu$m image, neither the source or companion are visible, yet the galaxy is clearly seen and lies within the {\it IRAS} 100 $\mu$m beam, centered over the source position. The SNR of the signal, returned with {\it SCANPI} at 60 $\mu$m is marginal at 2.9$\sigma$ and the positional offset from the source to signal peak is 0.33\arcmin. Whilst this offset is below the FSC criteria to flag the source as an upper limit, it is significantly greater than that of other 60 $\mu$m source detections (as determined through comparison of our sources and those of a similar study i.e. \citealp{Williams:2006} - as is discussed in Section 6). 

From the MIPS images we have measured the source flux at 24 $\mu$m, the 3$\sigma$ upper limit at 70 $\mu$m, and the flux of the galaxy (light blue point); included on the SED shown in Figure \ref{fig:SEDcontam}. We find that the source 24 $\mu$m flux lies on the modelled photosphere of this object and that the 70 $\mu$m upper limit lies significantly below the single blackbody, model excess fit. Furthermore, we find that the 70 $\mu$m flux of the galaxy matches well to the excess fit. In visual inspection of the {\it IRAS} scans it can be seen that at the position of the galaxy, the corresponding pixel becomes increasingly brighter at longer wavelengths. From this analysis, we conclude that both the {\it IRAS} 60 $\mu$m and 100 $\mu$m flux is contaminated due to the emission from the galaxy.

% ################# HIGHER RESOLUTION IR IMAGES  #################
% ################# END HIGHER RESOLUTION IR IMAGES  #################

% ################# SPITZER HIP112460  #################
\begin{figure}
\centering
\includegraphics[scale=0.28]{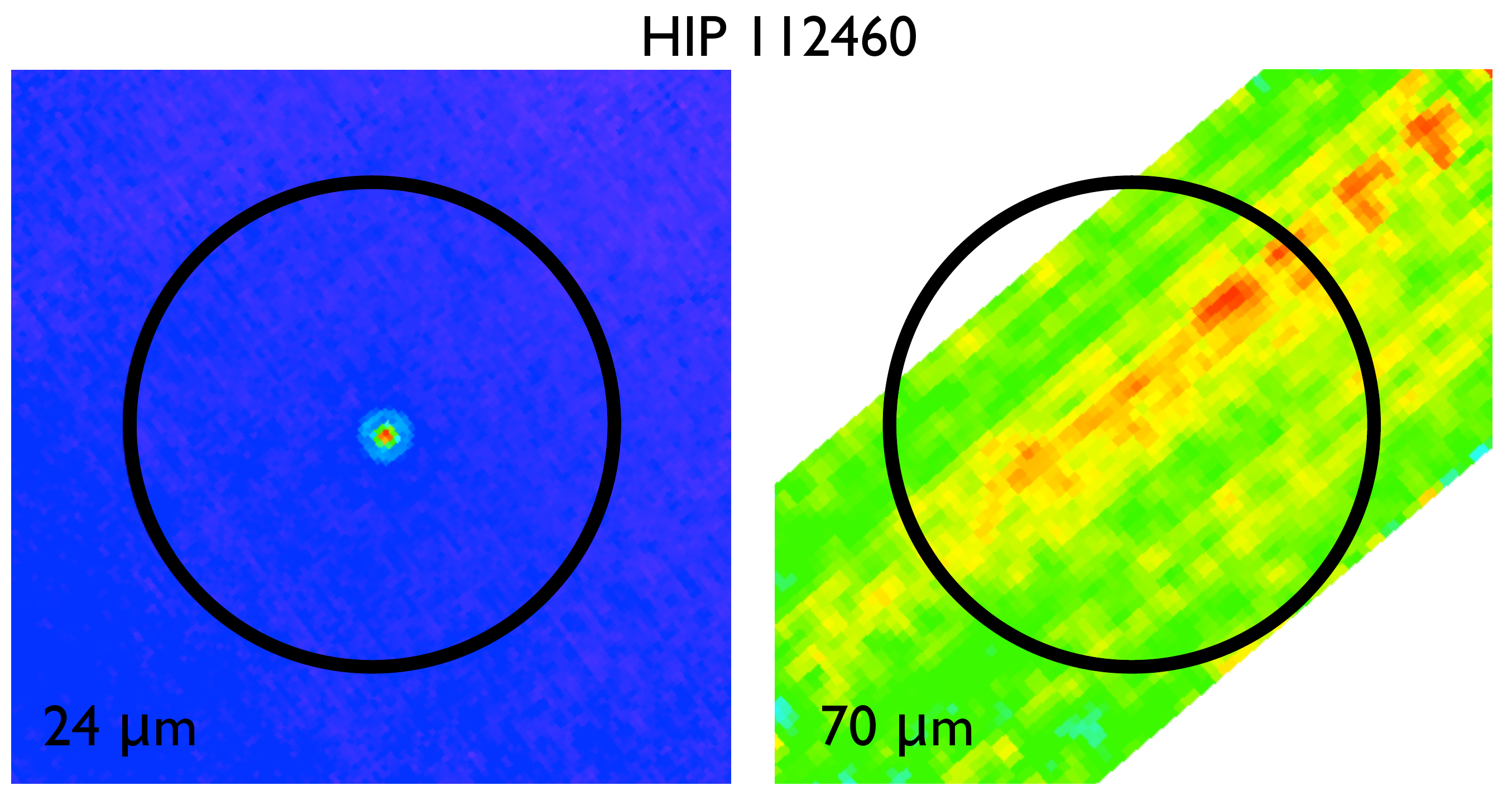}
\caption{Higher resolution {\it Spitzer} MIPS images for HIP 112460. The orientation of both images are North, up and East, left and the images are 300\arcsec in size along each axis. The black circles represent the {\it IRAS} 100~$\mu$m beam centered over the source position. The source is clearly detected in the 24 $\mu$m image, with no visible nearby contaminating objects. The 3$\sigma$ upper limit of the source measured in the 70 $\mu$m image lies well below the modelled, single blackbody excess fit. Whilst stripping artifacts are seen in the 70 $\mu$m image no nearby contaminating objects are visible.}
\label{fig:hip112460}
\end{figure}
% ################# END SPITZER HIP112460  #################

\subsubsection{HIP 112460}
Although the HIP 112460 M-star cold debris disk candidate has the largest discrepancy between the predicted submm flux from the SED and the measured CSO limit (a factor of $\sim$100), no additional source was identified within the {\it IRAS} beam. From inspection of the higher resolution {\it WISE} and MIPS images, no contaminating object is detected in the surrounding field of this source, as shown in Figure \ref{fig:hip112460}. Due to the low latitude of the source (b = -13\degr), it is possible that the associated 100 $\mu$m {\it IRAS} flux is contaminated from Galactic cirrus emission. Another possibility for this nearby source is a dilution of the flux from a disk due to a spatially resolved disk. To assess the impact of a spatially resolved disk, we searched for the largest reported spatially resolved disks around M-stars (AU Mic -- \citealp{MacGregor:2013} and GJ 581 -- \citealp{Lestrade:2012}), scaled the disk size based on the distance to HIP 112460, and determined the number of beams across which the flux would be distributed in a uniform disk. Even for the larger of the two resolved M-star disks, the resulting size at the distance of HIP 112460 would spread the flux across $\sim$9 beams, diluting the flux by a factor too small to account for the large discrepancy. 

Another possible explanation for the submm non-detection is a variable source. Some support for this hypothesis comes from the conflicting flux measurements at different epochs, since the fluxes obtained from later observations with {\it WISE} at 12 $\mu$m and 22 $\mu$m and {\it Spitzer} MIPS at 24 $\mu$m fall on the photosphere of the star, compared to the excesses obtained from the earlier {\it IRAS} at 12 and 25 $\mu$m scans, and seen at high significance at 12 $\mu$m and 100 $\mu$m in Figure \ref{fig:IRAS}. From the {\it Spitzer} MIPS images of this source, we measured the 24~$\mu$m flux and the 70~$\mu$m upper limit. As shown in Figure \ref{fig:SEDcontam}, the {\it Spitzer} 24 $\mu$m flux lies on the modelled stellar photosphere for this source and both the 3$\sigma$ upper limits at 70~$\mu$m and our 350 $\mu$m observation, lie well below the modelled, single blackbody excess fit (unlike the {\it IRAS} values). A large scale {\it IRAS} variability investigation of blazars \citep{Impey:1988} did find sources with variability up to an order of magnitude at 100 $\mu$m, however such a large amplitude change was very rare. Another example of a large degree of variability intrinsic to a debris disk at somewhat shorter wavelengths in the mid-IR was reported for the TYC 8241 2652 1 debris disk \citep{Melis:2012} which exhibited a remarkable change of a factor of $\sim$30 over a timescale of less than two years. More observations are required to provide further understanding into the nature of this unusual, possibly variable, system.

% ################# END SECTION: RESULTS ################# 

% ################# SECTION: INVESTIGATION OF OTHER DEBRIS DISKS #################
 
\section{Investigation of other debris disks with upper limits}
The high rate of source contamination of the {\it IRAS} FSC fluxes found amongst our cold debris disk sample prompted us to conduct the same contamination analysis for a previous submm study, at 850 $\mu$m of eight debris disks \citep{Williams:2006}, to further assess the likelihood of additional examples of misidentified debris disks. In the \citet{Williams:2006} study, the sources were selected through {\it IRAS} FSC excesses, similar to our program. Upper-limits were obtained for two of the eight sources -- HD 56099 and HD 78702. From inspection of the higher resolution {\it WISE} and {\it Spitzer} MIPS images of these two sources, we find that a nearby object lies within the {\it IRAS} beam size for HD 56099). No nearby contaminating objects are visible in either the {\it WISE} or {\it Spitzer} images for HD 78702. For HD~56099, the nearby object is visible in the {\it WISE} W3 and W4 images and in the MIPS 24 $\mu$m, 70 $\mu$m and 160 $\mu$m images, and becomes increasingly brighter with wavelength. \citet{Moor:2006} identify this source to be a misclassified debris disk system, due to the nearby object contamination as identified, and \citet{Moor:2011} have further analysed the {\it Spitzer} observations for this target and find a low significance level of excess above that of the stellar photosphere.
One out of eight sources, in the study carried out by \citet{Williams:2006}, was misidentified of possessing a cold debris disk. Comparatively, for the cold debris disk sample within this paper, three out of the five sources show evidence of nearby object contamination of the associated {\it IRAS} FSC fluxes.

% ################# END SECTION: INVESTIGATION OF OTHER DEBRIS DISKS ################# 

% ################# SECTION: CONCLUSIONS ################# 
\section{Conclusions}
We have presented new submm observations at either 350~$\mu$m or 450~$\mu$m, searching for dust emission around a sample of cold and warm nearby ($<$ 40pc) debris disk candidates that have been identified from {\it IRAS} or {\it WISE} excesses. 

Upper-limits were obtained for all six targets, and four of the limits are well below the expected submm flux, based on a single temperature blackbody model fit to previous shorter wavelength photometry. Of the four cold excess sources with very discrepant submm limits compared to the SED fit, we found that three targets have clearly identified contaminating objects in the {\it IRAS} beam, and, therefore, are likely misidentified debris disk candidates. Quantitive measurements of the contaminating objects were obtained for two targets --  HIP~51658 and HIP~76375 -- from higher resolution, pointed {\it Herschel} PACS and {\it Spitzer} MIPS observations at wavelengths comparable to the {\it IRAS} excess fluxes. From visual inspection of the {\it WISE} all-sky survey images, we find that HIP~61680 is also like to suffer from {\it IRAS} contamination, in which we identify a nearby, background object that becomes increasingly brighter over the {\it WISE} bands as the likely true source of the {\it IRAS} FSC measurement. 

The rate of {\it IRAS} contamination has likely lead to a greater number of sources in being identified as possessing cold debris disks than the actual number. Considering the high level of source contamination in this study and comparable results from an analysis of other debris disks candidates with submm non-detections \citep{Williams:2006}, lower detection rates than anticipated for large scale debris disks surveys that are currently being pursued with {\it Herschel} and SCUBA2 may be found. 

It is important to highlight the need for careful examination of source contamination for observations in the far-IR and submm wavelengths, where the probability of detecting a nearby background source can be as high as 36\% \citep{Booth:2013, Berta:2011}. In this study we find 4-5 examples of contamination either from identified objects or possible background variability, giving a contamination rate of 31-38\% similar to the probability reported in \citet{Booth:2013}.

Finally, we note that it is only now that it is possible to assess systematically the {\it IRAS} associations of debris disk candidates by using the complementary mid-IR to far-IR wavelength coverage and higher resolution and sensitivity provided by {\it WISE}, {\it Spitzer} and {\it Herschel}.

% ################# END SECTION: CONCLUSIONS ################# 

% #######################################################  
% ################# ACKNOWLEDGEMENTS #################
\begin{acknowledgements}
This material is based upon work at the Caltech Submillimeter Observatory, which is operated by the California Institute of Technology under cooperative agreement with the National Science Foundation (AST-0838261) We gratefully acknowledge support to Exeter from the Leverhulme Trust (F/00144/BJ) and STFC (ST/F0071241/1, ST/H002707/1). This research is partly based on data obtained from the ESO Science Archive Facility, and the CFHT Science Archive. Based on observations made with the NASA/ESA Hubble Space Telescope, obtained from the data archive at the Space Telescope Science Institute. STScI is operated by the Association of Universities for Research in Astronomy, Inc. under NASA contract NAS 5-26555. This research has made use of the SIMBAD database, operated at CDS, Strasbourg, France. This publication makes use of data products from the Two Micron All Sky Survey, which is a joint project of the University of Massachusetts and the Infrared Processing and Analysis Center/California Institute of Technology, funded by the National Aeronautics and Space Administration and the National Science Foundation. This research has made use of the Washington Double Star Catalog maintained at the U.S. Naval Observatory. This research has made use of data obtained from the SuperCOSMOS Science Archive, prepared and hosted by the Wide Field Astronomy Unit, Institute for Astronomy, University of Edinburgh, which is funded by the UK Science and Technology Facilities Council. This research used the facilities of the Canadian Astronomy Data Centre operated by the National Research Council of Canada with the support of the Canadian Space Agency. This research has made use of the NASA/IPAC Infrared Science Archive, which is operated by the Jet Propulsion Laboratory, California Institute of Technology, under contract with the National Aeronautics and Space Administration. This publication makes use of data products from the Wide-field Infrared Survey Explorer, which is a joint project of the University of California, Los Angeles, and the Jet Propulsion Laboratory/California Institute of Technology, funded by the National Aeronautics and Space Administration.
\end{acknowledgements}
% ################# REFERENCES #################
\bibliographystyle{aa}
%\bibliography{debris}

\begin{thebibliography}{71}
\expandafter\ifx\csname natexlab\endcsname\relax\def\natexlab#1{#1}\fi

\bibitem[{Aumann {et~al.}(1984)Aumann, Beichman, Gillett, de~Jong, Houck, Low,
  Neugebauer, Walker, \& Wesselius}]{Aumann:1984}
Aumann, H.~H., Beichman, C.~A., Gillett, F.~C., {et~al.} 1984, Astrophysical
  Journal, 278, L23

\bibitem[{Backman \& Paresce(1993)}]{Backman:1993}
Backman, D.~E. \& Paresce, F. 1993, In: Protostars and planets III (A93-42937
  17-90), 1253

\bibitem[{Balega {et~al.}(2007)Balega, Balega, Maksimov, Malogolovets,
  Rastegaev, Shkhagosheva, \& Weigelt}]{Balega:2007}
Balega, I.~I., Balega, Y.~Y., Maksimov, A.~F., {et~al.} 2007, Astrophysical
  Bulletin, 62, 339

\bibitem[{Baraffe {et~al.}(1998)Baraffe, Chabrier, Allard, \&
  Hauschildt}]{Baraffe:1998}
Baraffe, I., Chabrier, G., Allard, F., \& Hauschildt, P.~H. 1998, Astronomy and
  Astrophysics, 337, 403

\bibitem[{Beichman {et~al.}(2006)Beichman, Tanner, Bryden, Stapelfeldt, Werner,
  Rieke, Trilling, Lawler, \& Gautier}]{Beichman:2006}
Beichman, C.~A., Tanner, A., Bryden, G., {et~al.} 2006, The Astrophysical
  Journal, 639, 1166

\bibitem[{Berta {et~al.}(2011)Berta, Magnelli, Nordon, Lutz, Wuyts, Altieri,
  Andreani, Aussel, Casta{\~n}eda, Cepa, Cimatti, Daddi, Elbaz, Schreiber,
  Genzel, Floc'h, Maiolino, P{\'e}rez-Fournon, Poglitsch, Popesso, Pozzi,
  Riguccini, Rodighiero, Sanchez-Portal, Sturm, Tacconi, \&
  Valtchanov}]{Berta:2011}
Berta, S., Magnelli, B., Nordon, R., {et~al.} 2011, Astronomy {\&}
  Astrophysics, 532, 49

\bibitem[{Booth {et~al.}(2013)Booth, Kennedy, Sibthorpe, Matthews, Wyatt,
  Duch{\^e}ne, Kavelaars, Rodriguez, Greaves, Koning, Vican, Rieke, Su,
  Moro-Mart{\'\i}n, \& Kalas}]{Booth:2013}
Booth, M., Kennedy, G., Sibthorpe, B., {et~al.} 2013, Monthly Notices of the
  Royal Astronomical Society, 428, 1263

\bibitem[{Bryden {et~al.}(2006)Bryden, Beichman, Trilling, Rieke, Holmes,
  Lawler, Stapelfeldt, Werner, Gautier, Blaylock, Gordon, Stansberry, \&
  Su}]{Bryden:2006}
Bryden, G., Beichman, C.~A., Trilling, D.~E., {et~al.} 2006, The Astrophysical
  Journal, 636, 1098

\bibitem[{Cameron(1997)}]{Cameron:1997}
Cameron, A. G.~W. 1997, Icarus, 126, 126

\bibitem[{Carpenter {et~al.}(2009)Carpenter, Bouwman, Mamajek, Meyer,
  Hillenbrand, Backman, Henning, Hines, Hollenbach, Kim, Moro-Martin, Pascucci,
  Silverstone, Stauffer, \& Wolf}]{Carpenter:2009}
Carpenter, J.~M., Bouwman, J., Mamajek, E.~E., {et~al.} 2009, The Astrophysical
  Journal Supplement, 181, 197

\bibitem[{Cutri \& et~al.(2012)}]{Cutri:2012}
Cutri, R.~M. \& et~al. 2012, VizieR On-line Data Catalog, 2311, 0

\bibitem[{Dent {et~al.}(2000)Dent, Walker, Holland, \& Greaves}]{Dent:2000}
Dent, W. R.~F., Walker, H.~J., Holland, W.~S., \& Greaves, J.~S. 2000, Monthly
  Notices of the Royal Astronomical Society, 314, 702

\bibitem[{Dowell {et~al.}(2003)Dowell, Allen, Babu, Freund, Gardner, Groseth,
  Jhabvala, Kovacs, Lis, Moseley, Phillips, Silverberg, Voellmer, \&
  Yoshida}]{Dowell:2003}
Dowell, C.~D., Allen, C.~A., Babu, R.~S., {et~al.} 2003, Millimeter and
  Submillimeter Detectors for Astronomy. Edited by Phillips, 4855, 73

\bibitem[{Dunne \& Eales(2001)}]{Dunne:2001}
Dunne, L. \& Eales, S.~A. 2001, Monthly Notices of the Royal Astronomical
  Society, 327, 697

\bibitem[{Griffin \& Orton(1993)}]{Griffin:1993}
Griffin, M.~J. \& Orton, G.~S. 1993, Icarus, 105, 537

\bibitem[{Habing {et~al.}(1999)Habing, Dominik, de~Muizon, Kessler, Laureijs,
  Leech, Metcalfe, Salama, Siebenmorgen, \& Trams}]{Habing:1999}
Habing, H.~J., Dominik, C., de~Muizon, M.~J., {et~al.} 1999, Nature, 401, 456

\bibitem[{Hambly {et~al.}(2001)Hambly, MacGillivray, Read, Tritton, Thomson,
  Kelly, Morgan, Smith, Driver, Williamson, Parker, Hawkins, Williams, \&
  Lawrence}]{Hambly:2001}
Hambly, N.~C., MacGillivray, H.~T., Read, M.~A., {et~al.} 2001, Monthly Notices
  of the Royal Astronomical Society, 326, 1279

\bibitem[{Hartkopf \& Mason(2009)}]{Hartkopf:2009}
Hartkopf, W.~I. \& Mason, B.~D. 2009, The Astronomical Journal, 138, 813

\bibitem[{Hartkopf \& McAlister(1984)}]{Hartkopf:1984}
Hartkopf, W.~I. \& McAlister, H.~A. 1984, Astronomical Society of the Pacific,
  96, 105

\bibitem[{Hauschildt {et~al.}(1999)Hauschildt, Allard, \&
  Baron}]{Hauschildt:1999}
Hauschildt, P.~H., Allard, F., \& Baron, E. 1999, The Astrophysical Journal,
  512, 377

\bibitem[{Hillenbrand {et~al.}(2008)Hillenbrand, Carpenter, Kim, Meyer,
  Backman, Moro-Mart{\'\i}n, Hollenbach, Hines, Pascucci, \&
  Bouwman}]{Hillenbrand:2008}
Hillenbrand, L.~A., Carpenter, J.~M., Kim, J.~S., {et~al.} 2008, The
  Astrophysical Journal, 677, 630

\bibitem[{Holland {et~al.}(1998)Holland, Greaves, Zuckerman, Webb, McCarthy,
  Coulson, Walther, Dent, Gear, \& Robson}]{Holland:1998}
Holland, W.~S., Greaves, J.~S., Zuckerman, B., {et~al.} 1998, Nature, 392, 788

\bibitem[{Hughes {et~al.}(2011)Hughes, Wilner, Andrews, Williams, Su,
  Murray-Clay, \& Qi}]{Hughes:2011}
Hughes, A.~M., Wilner, D.~J., Andrews, S.~M., {et~al.} 2011, The Astrophysical
  Journal, 740, 38

\bibitem[{Impey \& Neugebauer(1988)}]{Impey:1988}
Impey, C.~D. \& Neugebauer, G. 1988, Astronomical Journal (ISSN 0004-6256), 95,
  307

\bibitem[{Jayawardhana {et~al.}(2002)Jayawardhana, Holland, Kalas, Greaves,
  Dent, Wyatt, \& Marcy}]{Jayawardhana:2002}
Jayawardhana, R., Holland, W.~S., Kalas, P., {et~al.} 2002, The Astrophysical
  Journal, 570, L93

\bibitem[{Jenness {et~al.}(2002)Jenness, Stevens, Archibald, Economou, Jessop,
  \& Robson}]{Jenness:2002}
Jenness, T., Stevens, J.~A., Archibald, E.~N., {et~al.} 2002, Monthly Notice of
  the Royal Astronomical Society, 336, 14

\bibitem[{Kov{\'a}cs(2006)}]{Kovacs:2006}
Kov{\'a}cs, A. 2006, Proquest Dissertations And Theses 2006. Section 0036, 28

\bibitem[{Kov{\'a}cs(2008)}]{Kovacs:2008}
Kov{\'a}cs, A. 2008, Millimeter and Submillimeter Detectors and Instrumentation
  for Astronomy IV. Edited by Duncan, 7020, 45

\bibitem[{Kuchner \& Holman(2003)}]{Kuchner:2003}
Kuchner, M.~J. \& Holman, M.~J. 2003, The Astrophysical Journal, 588, 1110

\bibitem[{Leong {et~al.}(2006)Leong, Peng, Houde, Yoshida, Chamberlin, \&
  Phillips}]{Leong:2006}
Leong, M., Peng, R., Houde, M., {et~al.} 2006, Millimeter and Submillimeter
  Detectors and Instrumentation for Astronomy III. Edited by Zmuidzinas, 6275,
  21

\bibitem[{Lestrade {et~al.}(2012)Lestrade, Matthews, Sibthorpe, Kennedy, Wyatt,
  Bryden, Greaves, Thilliez, Moro-Mart{\'\i}n, Booth, Dent, Duch{\^e}ne,
  Harvey, Horner, Kalas, Kavelaars, Phillips, Rodriguez, Su, \&
  Wilner}]{Lestrade:2012}
Lestrade, J.-F., Matthews, B.~C., Sibthorpe, B., {et~al.} 2012, Astronomy {\&}
  Astrophysics, 548, 86

\bibitem[{Lestrade {et~al.}(2006)Lestrade, Wyatt, Bertoldi, Dent, \&
  Menten}]{Lestrade:2006}
Lestrade, J.-F., Wyatt, M.~C., Bertoldi, F., Dent, W. R.~F., \& Menten, K.~M.
  2006, Astronomy and Astrophysics, 460, 733

\bibitem[{Liou \& Zook(1999)}]{Liou:1999}
Liou, J.-C. \& Zook, H.~A. 1999, The Astronomical Journal, 118, 580

\bibitem[{Lisse {et~al.}(2002)Lisse, Schultz, Fernandez, Peschke, Ressler,
  Gorjian, Djorgovski, Christian, Golisch, \& Kaminski}]{Lisse:2002}
Lisse, C., Schultz, A., Fernandez, Y., {et~al.} 2002, The Astrophysical
  Journal, 570, 779

\bibitem[{Liu {et~al.}(2004)Liu, Matthews, Williams, \& Kalas}]{Liu:2004}
Liu, M.~C., Matthews, B.~C., Williams, J.~P., \& Kalas, P.~G. 2004, The
  Astrophysical Journal, 608, 526

\bibitem[{MacGregor {et~al.}(2013)MacGregor, Wilner, Rosenfeld, Andrews,
  Matthews, Hughes, Booth, Chiang, Graham, Kalas, Kennedy, \&
  Sibthorpe}]{MacGregor:2013}
MacGregor, M.~A., Wilner, D.~J., Rosenfeld, K.~A., {et~al.} 2013, The
  Astrophysical Journal Letters, 762, L21

\bibitem[{Mannings \& Barlow(1998)}]{Mannings:1998}
Mannings, V. \& Barlow, M.~J. 1998, Astrophysical Journal v.497, 497, 330

\bibitem[{McCarthy {et~al.}(1998)McCarthy, Ge, Hinz, Finn, Low, Cheselka, \&
  Salvestrini}]{McCarthy:1998}
McCarthy, D.~W., Ge, J., Hinz, J.~L., {et~al.} 1998, American Astronomical
  Society, 193, 1265

\bibitem[{Melis {et~al.}(2012)Melis, Zuckerman, Rhee, Song, Murphy, \&
  Bessell}]{Melis:2012}
Melis, C., Zuckerman, B., Rhee, J.~H., {et~al.} 2012, Nature, 487, 74

\bibitem[{Miura {et~al.}(1995)Miura, Iribe, Kubo, Baba, \& Isobe}]{Miura:1995}
Miura, N., Iribe, T., Kubo, T., Baba, N., \& Isobe, S. 1995, Publications of
  the National Astronomical Observatory of Japan, 4, 67

\bibitem[{Mo{\'o}r {et~al.}(2006)Mo{\'o}r, {\'A}brah{\'a}m, Derekas, Kiss,
  Kiss, Apai, Grady, \& Henning}]{Moor:2006}
Mo{\'o}r, A., {\'A}brah{\'a}m, P., Derekas, A., {et~al.} 2006, The
  Astrophysical Journal, 644, 525

\bibitem[{Mo{\'o}r {et~al.}(2011)Mo{\'o}r, Pascucci, K{\'o}sp{\'a}l,
  {\'A}brah{\'a}m, Csengeri, Kiss, Apai, Grady, Henning, Kiss, Bayliss,
  Juh{\'a}sz, Kov{\'a}cs, \& Szalai}]{Moor:2011}
Mo{\'o}r, A., Pascucci, I., K{\'o}sp{\'a}l, {\'A}., {et~al.} 2011, The
  Astrophysical Journal Supplement, 193, 4

\bibitem[{Moshir {et~al.}(1992)Moshir, Kopman, \& Conrow}]{Moshir:1992}
Moshir, M., Kopman, G., \& Conrow, T. A.~O. 1992, Pasadena: Infrared Processing
  and Analysis Center

\bibitem[{Najita \& Williams(2005)}]{Najita:2005}
Najita, J. \& Williams, J.~P. 2005, The Astrophysical Journal, 635, 625

\bibitem[{Patience {et~al.}(2011)Patience, Bulger, King, Ayliffe, Bate, Song,
  Pinte, Koda, Dowell, \& Kov{\'a}cs}]{Patience:2011}
Patience, J., Bulger, J., King, R.~R., {et~al.} 2011, Astronomy {\&}
  Astrophysics, 531, L17

\bibitem[{Perryman \& ESA(1997)}]{Perryman:1997}
Perryman, M. A.~C. \& ESA. 1997, The Hipparcos and Tycho catalogues.
  Astrometric and photometric star catalogues derived from the ESA Hipparcos
  Space Astrometry Mission, 1200, iSBN: 9290923997

\bibitem[{Poglitsch {et~al.}(2010)Poglitsch, Waelkens, Geis, Feuchtgruber,
  Vandenbussche, Rodriguez, Krause, Renotte, van Hoof, Saraceno, Cepa,
  Kerschbaum, Agn{\`e}se, Ali, Altieri, Andreani, Augueres, Balog, Barl, Bauer,
  Belbachir, Benedettini, Billot, Boulade, Bischof, Blommaert, Callut, Cara,
  Cerulli, Cesarsky, Contursi, Creten, Meester, Doublier, Doumayrou, Duband,
  Exter, Genzel, Gillis, Gr{\"o}zinger, Henning, Herreros, Huygen, Inguscio,
  Jakob, Jamar, Jean, de~Jong, Katterloher, Kiss, Klaas, Lemke, Lutz, Madden,
  Marquet, Martignac, Mazy, Merken, Montfort, Morbidelli, M{\"u}ller, Nielbock,
  Okumura, Orfei, Ottensamer, Pezzuto, Popesso, Putzeys, Regibo, Reveret,
  Royer, Sauvage, Schreiber, Stegmaier, Schmitt, Schubert, Sturm, Thiel,
  Tofani, Vavrek, Wetzstein, Wieprecht, \& Wiezorrek}]{Poglitsch:2010}
Poglitsch, A., Waelkens, C., Geis, N., {et~al.} 2010, Astronomy and
  Astrophysics, 518, L2

\bibitem[{Pollack {et~al.}(1994)Pollack, Hollenbach, Beckwith, Simonelli,
  Roush, \& Fong}]{Pollack:1994}
Pollack, J.~B., Hollenbach, D., Beckwith, S., {et~al.} 1994, Astrophysical
  Journal, 421, 615

\bibitem[{Quillen \& Faber(2006)}]{Quillen:2006}
Quillen, A.~C. \& Faber, P. 2006, Monthly Notices of the Royal Astronomical
  Society, 373, 1245

\bibitem[{Rhee {et~al.}(2007)Rhee, Song, Zuckerman, \& McElwain}]{Rhee:2007}
Rhee, J.~H., Song, I., Zuckerman, B., \& McElwain, M. 2007, The Astrophysical
  Journal, 660, 1556

\bibitem[{Rieke {et~al.}(2005)Rieke, Su, Stansberry, Trilling, Bryden,
  Muzerolle, White, Gorlova, Young, Beichman, Stapelfeldt, \&
  Hines}]{Rieke:2005}
Rieke, G.~H., Su, K. Y.~L., Stansberry, J.~A., {et~al.} 2005, The Astrophysical
  Journal, 620, 1010

\bibitem[{Roeser {et~al.}(2010)Roeser, Demleitner, \& Schilbach}]{Roeser:2010}
Roeser, S., Demleitner, M., \& Schilbach, E. 2010, The Astronomical Journal,
  139, 2440

\bibitem[{Sandell(2003)}]{Sandell:2003}
Sandell, G. 2003, `The calibration legacy of the ISO Mission', 481, 439

\bibitem[{Shaya \& Olling(2011)}]{Shaya:2011}
Shaya, E.~J. \& Olling, R.~P. 2011, The Astrophysical Journal Supplement, 192,
  2

\bibitem[{Sheret {et~al.}(2004)Sheret, Dent, \& Wyatt}]{Sheret:2004}
Sheret, I., Dent, W. R.~F., \& Wyatt, M.~C. 2004, Monthly Notices of the Royal
  Astronomical Society, 348, 1282

\bibitem[{Silverstone(2000)}]{Silverstone:2000}
Silverstone, M.~D. 2000, Thesis (PhD). UNIVERSITY OF CALIFORNIA, 17

\bibitem[{Smith \& Terrile(1984)}]{Smith:1984}
Smith, B.~A. \& Terrile, R.~J. 1984, Science (ISSN 0036-8075), 226, 1421

\bibitem[{Song {et~al.}(2002)Song, Weinberger, Becklin, Zuckerman, \&
  Chen}]{Song:2002}
Song, I., Weinberger, A.~J., Becklin, E.~E., Zuckerman, B., \& Chen, C. 2002,
  The Astronomical Journal, 124, 514

\bibitem[{Stern(1996)}]{Stern:1996}
Stern, S.~A. 1996, Astronomy and Astrophysics, 310, 999

\bibitem[{Su {et~al.}(2006)Su, Rieke, Stansberry, Bryden, Stapelfeldt,
  Trilling, Muzerolle, Beichman, Moro-Martin, Hines, \& Werner}]{Su:2006}
Su, K. Y.~L., Rieke, G.~H., Stansberry, J.~A., {et~al.} 2006, The Astrophysical
  Journal, 653, 675

\bibitem[{Sylvester {et~al.}(2001)Sylvester, Dunkin, \&
  Barlow}]{Sylvester:2001}
Sylvester, R.~J., Dunkin, S.~K., \& Barlow, M.~J. 2001, Monthly Notices of the
  Royal Astronomical Society, 327, 133

\bibitem[{van Belle(2010)}]{vanBelle:2010}
van Belle, G.~T. 2010, The Interferometric View on Hot Stars (Eds. Th. Rivinius
  {\&} M. Cur{\'e}) Revista Mexicana de Astronom{\'\i}a y Astrof{\'\i}sica
  (Serie de Conferencias) Vol. 38, 38, 119

\bibitem[{van Belle {et~al.}(2008)van Belle, van Belle, Creech-Eakman, Coyne,
  Boden, Akeson, Ciardi, Rykoski, Thompson, Lane, \&
  Collaboration}]{vanBelle:2008}
van Belle, G.~T., van Belle, G., Creech-Eakman, M.~J., {et~al.} 2008, The
  Astrophysical Journal Supplement Series, 176, 276

\bibitem[{van Leeuwen(2007)}]{vanLeeuwen:2007}
van Leeuwen, F. 2007, Astronomy and Astrophysics, 474, 653

\bibitem[{Williams \& Andrews(2006)}]{Williams:2006}
Williams, J.~P. \& Andrews, S.~M. 2006, The Astrophysical Journal, 653, 1480

\bibitem[{Wyatt(2006)}]{Wyatt:2006}
Wyatt, M.~C. 2006, The Astrophysical Journal, 639, 1153

\bibitem[{Wyatt(2008)}]{Wyatt:2008}
Wyatt, M.~C. 2008, Annual Review of Astronomy {\&} Astrophysics, 46, 339

\bibitem[{Wyatt {et~al.}(2003)Wyatt, Dent, \& Greaves}]{Wyatt:2003}
Wyatt, M.~C., Dent, W. R.~F., \& Greaves, J.~S. 2003, Monthly Notice of the
  Royal Astronomical Society, 342, 876

\bibitem[{Zacharias {et~al.}(2012)Zacharias, Finch, Girard, Henden, Bartlett,
  Monet, \& Zacharias}]{Zacharias:2012}
Zacharias, N., Finch, C.~T., Girard, T.~M., {et~al.} 2012, VizieR On-line Data
  Catalog, 1322, 0

\bibitem[{Zuckerman(2001)}]{Zuckerman:2001}
Zuckerman, B. 2001, Annual Review of Astronomy and Astrophysics, 39, 549

\bibitem[{Zuckerman \& Becklin(1993)}]{Zuckerman:1993}
Zuckerman, B. \& Becklin, E.~E. 1993, Astrophysical Journal, 414, 793

\end{thebibliography}

% ################# END DOCUMENT #################
\end{document}